\documentclass[twocolumn]{aastex63}

\usepackage{color}
\usepackage{multirow}
\usepackage{rotating}
\usepackage{longtable}
\usepackage{cancel,soul,ulem,amsmath} 

\def\hi{H{\sc i}}

\def\deg{$^\circ$}
\def\cm2{cm$^{-2}$}
\def\cc{cm$^{-3}$}
\def\kms{km s$^{-1}$}
\def\s{s$^{-1}$}
\def\nh3{NH$_3$}
\def\n2h{N$_2$H$^+$}

\def\13co{$^{13}$CO}
\def\c18o{C$^{18}$O}
\def\hc3n{HC$_3$N}
\def\h2{H$_2$}
\def\nh{n(H$_2$)}
\def\c2{[C\,{\sc ii}]}

\def\lc{\>\> ,}

\shorttitle{Precise Measurements of CH Maser Emission and Its Abundance in Translucent Clouds}
\shortauthors{Tang et al.}
\graphicspath{{./}{fig/}}

\begin{document}

\title{Precise Measurements of CH Maser Emission and  Its Abundance in Translucent Clouds}

\correspondingauthor{Ningyu Tang, Di Li, Gan Luo}
\email{nytang@ahnu.edu.cn, dili@nao.cas.cn, luogan@nju.edu.cn}

\author[0000-0002-2169-0472]{Ningyu Tang}
\affiliation{Department of Physics, Anhui Normal University, Wuhu, Anhui 241002, China}
\affiliation{National Astronomical Observatories, CAS, Beijing 100012, China}

\author[0000-0003-3010-7661]{Di Li}
\affiliation{National Astronomical Observatories, CAS, Beijing 100012, China}
\affiliation{NAOC-UKZN Computational Astrophysics Centre, University of KwaZulu-Natal, Durban 4000, South Africa}

\author[0000-0002-1583-8514]{Gan Luo}
\affiliation{School of Astronomy and Space Science, Nanjing University, 163 Xianlin Road, Nanjing 210023, China}

\author{ Carl Heiles}
\affiliation{ Department of Astronomy, University of California, Berkeley, 601 Campbell Hall 3411, Berkeley, CA 94720-3411, USA}

\author[0000-0003-2302-0613]{Sheng-Li Qin}
\affiliation{ Department of Astronomy, Yunnan University, Kunming 650091, China}

\author[0000-0001-6106-1171]{Junzhi Wang}
\affiliation{ Shanghai Astronomical Observatories, CAS}

\author{Jifeng Xia}
\affiliation{National Astronomical Observatories, CAS, Beijing 100012, China}

\author{Longfei Chen}
\affiliation{National Astronomical Observatories, CAS, Beijing 100012, China}



\begin{abstract}
We present high-sensitivity CH 9\,cm ON/OFF observations toward 18 extra-galactic continuum sources that have been detected with OH 18\,cm absorption in the Millennium survey with the Arecibo telescope. CH emission was detected toward six of eighteen sources. The excitation temperature of CH has been derived directly through analyzing all  detected ON and OFF velocity components. The excitation temperature of CH 3335 MHz transition ranges from $-54.5$ to $-0.4$ K and roughly follows a log-normal distribution peaking within [$-$5, 0] K, which implies overestimation by 20\% to more than ten times during calculating CH column density by assuming the conventional value of $-60$ or $-10$ K. Furthermore, the column density of CH would be underestimated by a factor of $1.32\pm 0.03$ when adopting  local thermal equilibrium (LTE) assumption instead of using the CH three hyperfine transitions. We found a  correlation between the  column density of CH and OH following log$N$(CH) = (1.80$\pm$ 0.49) log$N$(OH) $-11.59 \pm 6.87$.  The linear correlation between the column density of CH and \h2\ is consistent with that derived from visible wavelengths studies, confirming  that CH is one of the best tracers of \h2\ component in diffuse molecular gas.    
\end{abstract}



\section{Introduction} \label{sec:intro}

The CH radical, a simple hydride, is the first molecule detected in the interstellar medium (ISM) through the visible A$^2\Delta-$X$^2\Pi$ 4300-\r{A} absorption \citep{1937PASP...49...26D, 1937ApJ....86..483S, 1940PASP...52..187M}.  Absorption observations toward early-type stars in the visible wavelength (4300-\r{A}) show the ubiquitous existence of CH in the ISM  \citep[e.g.,][]{1982ApJ...257..125F,  1984A&A...130...62D, 2008ApJ...687.1075S}, and the visible 4300-\r{A} line  has been used extensively to identify the foreground cold gas components at a spectral resolution of 0.3 to 0.6 \kms \citep[e.g.,][]{1995MNRAS.277..458C, 1995ApJS...99..107C}.  The lack of bright, early-type background star is a major limitation on the usage of the 4300-\r{A} line \citep{2002A&A...391..693L}.

\begin{figure}
\begin{center}  
  \includegraphics[width=0.54\textwidth]{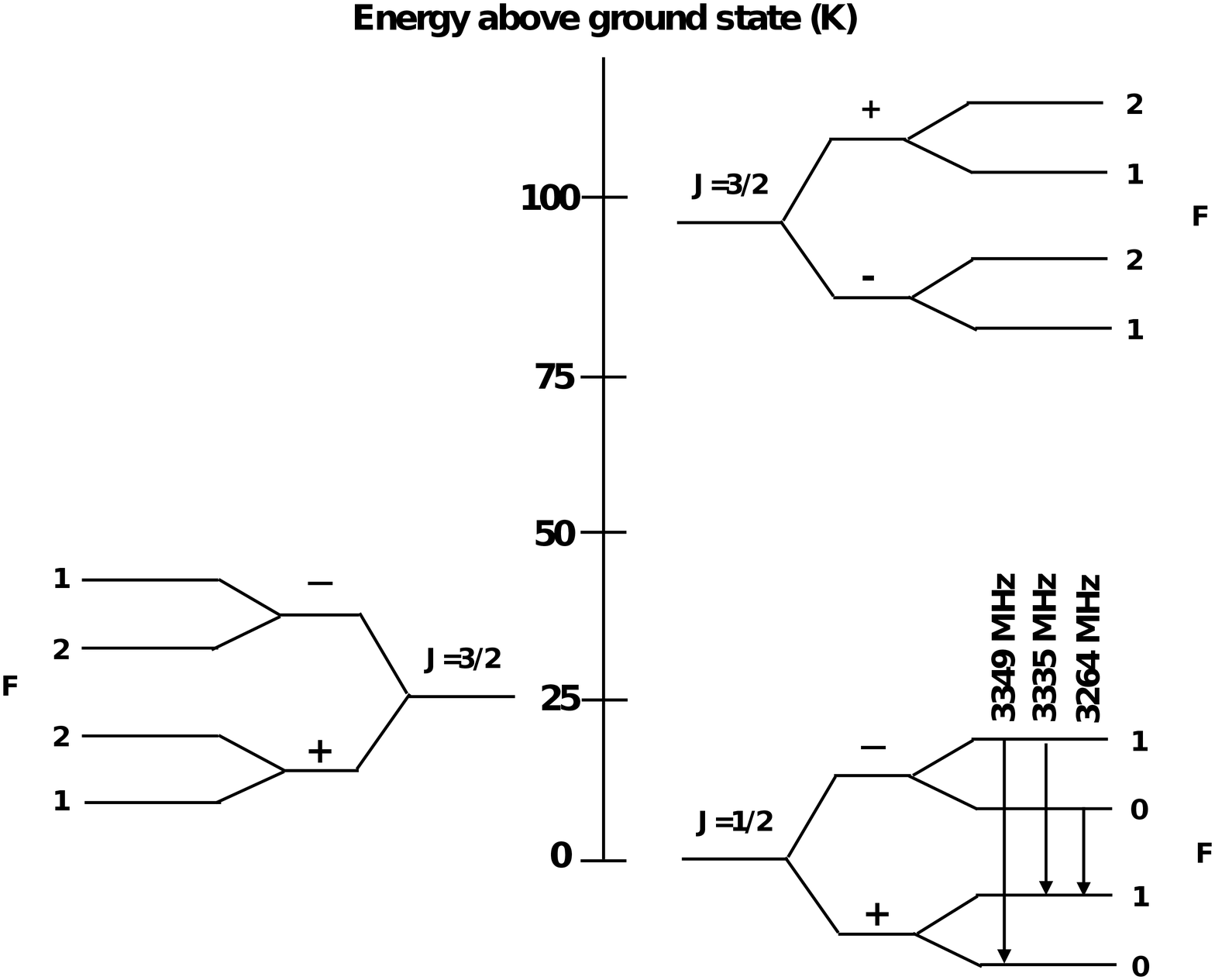}
\caption{ Energy-level diagram of 3 lowest  CH $^{2}\Pi$  rotational transitions. The 3335, 3264, and 3349 MHz transitions happen between different hyperfine $F$ levels of the state $^{2}\Pi_{1/2}, J=1/2$.  LTE intensity ratios between these transitions are $I$(3335):$I$(3264):$I$(3349)=2:1:1. The spacing of the $\Lambda$-doubled and hyperfine levels are not to scale. }
\label{fig:CH_energy}
\end{center}
\end{figure}

The $^2\Pi_{1/2}$, J=1/2 $\Lambda$-doubling radio  lines of CH (3264,  3335 and 3349 MHz; see Figure \ref{fig:CH_energy} for details) have been widely used in studying CH abundance and ISM evolution. In the past decades, radio studies reveal that the 3.3 GHz (9 cm) emissions exist in various star-forming environments   such as diffuse and translucent cloud \citep[e.g.,][]{1978ApJ...224..125L, 1980A&A....83..226S, 1981A&A....97..317S, 1986A&A...160..157M,  1993ApJ...408..559M, 2012A&A...546A.103S}, dark cloud \citep[e.g.,][]{1987A&A...173..347J}, and molecular outflows \citep[e.g.,][]{ 1988ApJ...329..920S, 1992A&AS...93..509M}.  

Chemically, CH is the first generation molecule in the carbon chemical network. Starting from $\rm C^{+}$ (e.g., Federman et al.\ 1984), 
\begin{equation}
\mathrm{C^+ + H_2 \rightarrow CH^+_2 + h\nu},
\end{equation}
CH is then synthesized by the following reactions
\begin{align} 
\rm CH_2^+ + e  & \rightarrow  \rm CH + H    \lc \nonumber \\ 
\rm CH_2^+ + H_2 & \rightarrow \rm CH_3^+ + H   \lc \nonumber \\
\rm CH_3^+ + e  & \rightarrow \rm CH + H_2\ or\ CH_2 + H .
\end{align}

Similar to CH,  hydroxyl (OH) is the first generation molecule in the oxygen chemical network which initiates from $\rm O^+$. The reactions could happen as follows 
\begin{align}
\rm O^+ + H_2 & \rightarrow \rm OH^+ + H \lc \nonumber \\
\rm OH^+ + H_2 & \rightarrow \rm H_2O^+ + H  \lc \nonumber \\
\rm H_2O^+ + H_2 & \rightarrow \rm H_3O^+ +H .
\end{align}
$\rm H_2O^+$ and $\rm H_3O^+$ then recombine with ambient electrons to form OH. 

Based on the above facts, both CH and OH could be adequate for tracing H$_2$, particularly in the HI-H$_2$ transition zone (Li et al.\ 2015, Xu \& Li 2016).  Noted that $\rm C^+$ is produced by the photoionization of atomic carbon,  while the $\rm O^+$ is produced through charge exchange between O and H$^+$, in which H$^+$ originates from cosmic ray ionization of H.

 The emission of CH and OH are generally weak (usually tens of mK in diffuse and translucent cloud), which requires a long integration time. Furthermore, the derivation of column density (therefore abundance) requires an exact value of excitation temperature and optical depth.  The combination of ON (absorption) and OFF (emission) observations toward continuum source provides direct measurements of these two parameters. This method has been widely applied for quantifying physical properties of \hi\  \citep[e.g.,][]{1983ApJS...53..591D}, OH \citep[e.g.,][]{1981A&A....98..271D}, and HCO$^+$ \citep[e.g.,][]{1996A&A...307..237L, 2020ApJ...889L...4L}.

Efforts have been taken to directly quantify the excitation of  CH through ON/OFF observations toward continuum sources since the 1970s.  A series of Galactic continuum sources (e.g.,  Cas A$^*$ ) and extra Galactic sources (e.g., 3C123) were surveyed with the Onsala Space Observatory 25.6 m telescope  \citep{1976ApJS...31..333R, 1977ApJS...35..263H}  and Effelsberg 100-m telescope \citep{1979A&A....73..253G}. Results based on these observations reveal that the $T\rm_{ex}$ of CH is inverted and ranges from $-60$ to $-9$ K but with relatively large uncertainty. The $T\rm_{ex}$ value of $-10$ K or $-60$ K derived from these observations is usually adopted to calculate the CH column density from 3335 MHz emission. This assumption may cause large deviation of calculating CH column density if $|T\rm_{ex}| < 3$ K  \citep{2020MNRAS.495..510D}. Recent modeling of CH 9 cm and 149 $\mu$m data toward four high-mass star-forming regions imply an extreme value, $T\rm_{ex} \sim -0.3$ K \citep{2021A&A...650A.133J}, which would lead to deviation of CH column density by a factor of 8.

 To further investigate  $T\rm_{ex}$ of CH in high sensitivity and to investigate the relationship between CH and OH in tracing \h2, sensitive CH and OH ON/OFF  spectra toward continuum sources are necessary. The Millennium survey took a high sensitive \hi\ absorption survey toward 79 continuum sources with Arecibo telescope \citep{ 2003ApJS..145..329H, 2003ApJ...586.1067H}.   Analysis results of corresponding OH data of this survey are present in  \citet{2018ApJS..235....1L}  and  \citet{2018ApJ...862...49N}. In this paper, we describe our results of a follow-up  CH survey toward 18 sightlines with detected OH absorption using the Arecibo telescope. 

The paper is organized as follows.  The observations and analysis are described in section \ref{sec:obs} and  \ref{sec:analysis}, respectively.  The derived CH properties are shown in section \ref{sec:CH_properties}.  We described the relationship between CH and OH in section \ref{subsec:ch_oh} and  the relationship between CH and \h2\  in section \ref{subsec:XCH}. Summary is presented in section \ref{sec:summary}.

\section{Observations}
\label{sec:obs}

\begin{figure*}
\begin{center}  
  \includegraphics[width=0.95\textwidth]{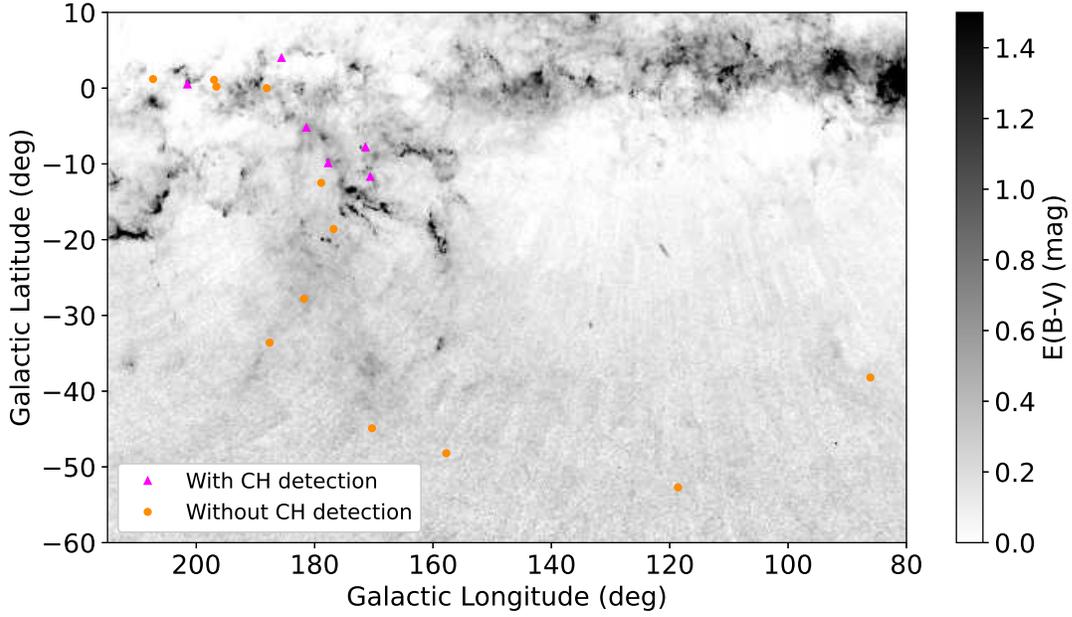}
\caption{ Spatial distribution of observed 18 sources overlaid on the Planck extinction map. The value of  color excess  E(B-V) ranges from 0 (white) to 1.5 (black) mag. }
\label{fig:src_distri} 
\end{center}
\end{figure*}

We have chosen 18 sources in the Millennium survey that have both \hi\ and OH absorption detections \citep{2018ApJS..235....1L}.  Figure \ref{fig:src_distri} shows the spatial distribution of the observed sources overlaid on Planck extinction map. 

Three $\Lambda$-doubling lines of CH in the $^2\Pi_{1/2}$, J=1/2 ground state were obtained with the Arecibo S-band receiver using the Wideband Arecibo Pulsar Processor (WAPP) backend in 2016 and 2019. The beam size (full width at half maximum, FWHM) is $\sim$1.5 arcmin at 3.3 GHz. 
The rest frequencies are 3335.481 MHz (F=1-1) for the main line,  3263.794 MHz (F=0-1) and 3349.193 MHz (F=1-0) for the two satellite lines.  Each spectrum has a bandwidth of 3.125 MHz in 8192 channels, resulting in a velocity resolution of 0.034 \kms. The data were then smoothed into 0.068 \kms\ to be comparable with OH data in \citet{2018ApJS..235....1L}.

During April and May 2016, observations were taken toward 18 sources and one OFF position (2 arcmin offset from the source in the north). The integration time is 10 minutes toward both ON and OFF  positions. The system temperature is $\sim$ 28 K, resulting in a reduced root-mean-square (rms)  of  $\sim$ 28 mK (antenna temperature) in the  OFF spectrum.  CH lines were detected toward six sources, namely 3C123, 3C131, 3C133, 3C154, T0526+24, and T0629+10.  During February and June 2019, we were able to observe  5 (3C123, 3C131, 3C133, 3C154, and T0629+10) of the above  6 sources with an ON position and 4 OFF positions (2 arcmin offset in the East, North, West, and South from the sources) in higher sensitivity. Twelve and three cycles of five-minute scans were taken for ON and each OFF position, respectively. This leads to a final rms of $\sim $ 9 mK in the OFF spectrum. 

\section{Analysis}
\label{sec:analysis}

\subsection{Radiation Transfer }
\label{subsec:radiation_transfer}

With the assumption of invariant CH distribution toward  ON and OFF positions, the obtained antenna temperature  $T\rm_A(\nu)$  after removing  continuum baseline for ON/OFF observations  are described  by 

\begin{equation}
T\rm_{A}^{ON}(\nu) =  \eta_b\eta_f(T_{ex}- T_{bg}-\eta_{cf} T_C^b )(1-e^{-\tau_{\nu}}) 
\label{eq:onspec}
\end{equation}

\begin{equation}
T\rm_{A}^{OFF}(\nu) =   \eta_b \eta_f(T_{ex}- T_{bg} )(1-e^{-\tau_{\nu}}) 
\label{eq:offspec}
\end{equation}
in which  $T\rm_{ex}$  and $\tau_{\nu}$ are the excitation temperature and optical depth of the foreground molecular cloud, respectively.   The brightness continuum temperature  $T_{bg}$ includes the contribution of both CMB (2.73 K) and Galactic synchrotron background, which is estimated as $\sim 0.07$ K at 3.3 GHz by applying spectral index of $-2.8$ from the 408 MHz survey  (Haslam et al.\ 1982).  $\eta_b$  is the main beam efficiency. It is 0.57$\pm$ 0.02 and 0.40 $\pm$ 0.04 before and after 2017 September 20, when the Arecibo telescope was hit by a hurricane.  $\eta_f$ is the beam filling factor of the cloud and is  assumed as 1.     

$\eta_{cf}$ and $T\rm_{C}^b$ are the  continuum source filling factor of the beam and brightness temperature of the background continuum sources, respectively.  For point sources,  $\eta_{b}\eta_{cf} T\rm_C^b$ equals $T\rm_C^A$, which is the antenna temperature of the point continuum source. 

For the sources with 4 OFF positions, their OFF spectra were obtained by averaging the spectra of 4 OFF positions. By combining the equation \ref{eq:onspec} and \ref{eq:offspec}, the excitation temperature $T\rm_{ex}$ and optical depth profile  $e^{-\tau_{\nu}}$ can be determined.  

\subsection{Spectral Line Fitting}
\label{subsec:fitline}

Following the method in \citet{2003ApJS..145..329H}, each profile was fitted with multiple Gaussian components to reach minimum variance. Firstly, we estimate the rms of each spectrum in the emission free velocity range (e.g., $-10$ to $-5$ \kms\ along 3C123). An effective decomposed component was generally certificated by the 3$\sigma$ criteria except  for the $-2.2$ \kms\ velocity component of the 3335 OFF spectrum toward 3C154. There are 3 cases according to the strength of CH intensity.

\begin{enumerate}
\item  CH emission is detected in both the ON and OFF spectrum and has a significant S/N ratio in the ON-OFF spectrum.  This is the case for the 3335\,MHz spectra of 3C123, 3C133, and T0629+10. For these sources, we first applied Gaussian decomposition to the optical depth profile, then the fitted central velocities of all components were fixed as initial values of the ON and OFF spectral fitting. 

\item CH emission is detected in both the ON and OFF spectrum but without  significant detection (3$\sigma$ level) in the ON-OFF spectrum. This is the case for the 3335\,MHz spectra of 3C131 and T0526+24. For these sources, we applied Gaussian decomposition to the ON profile, then the fitted central velocities were fixed as initial values of the OFF spectral fitting. 

\item CH emission is detected in the ON spectrum and is absent in the OFF spectrum. This is the case for the 3335\,MHz spectra of the 3C154 component with a center velocity of  $-1.3$ \kms.  Gaussian decomposition was adopted to the ON spectrum only.  

\end{enumerate}

\begin{figure*}
\gridline{\fig{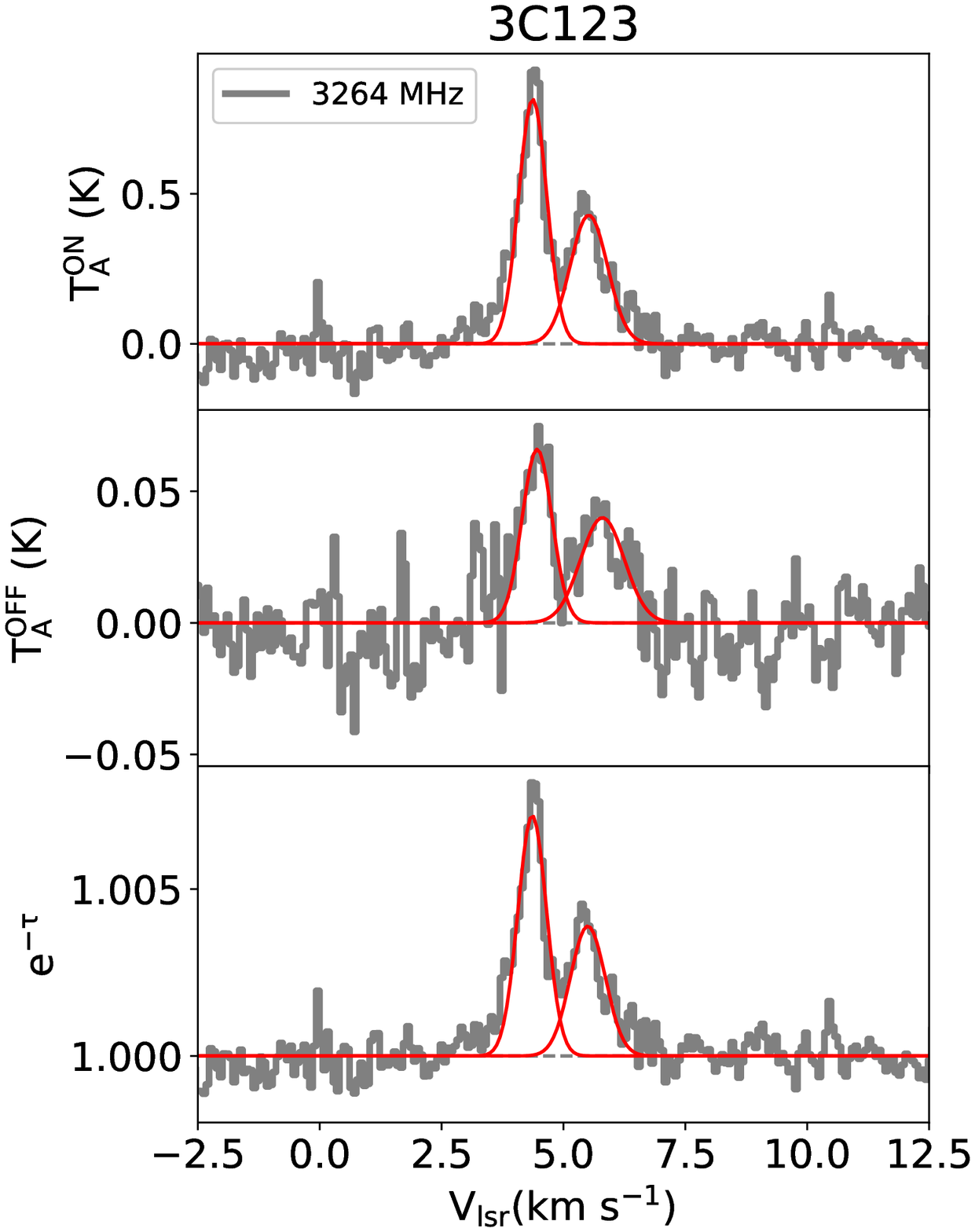}{0.32\textwidth}{(a)}
              \fig{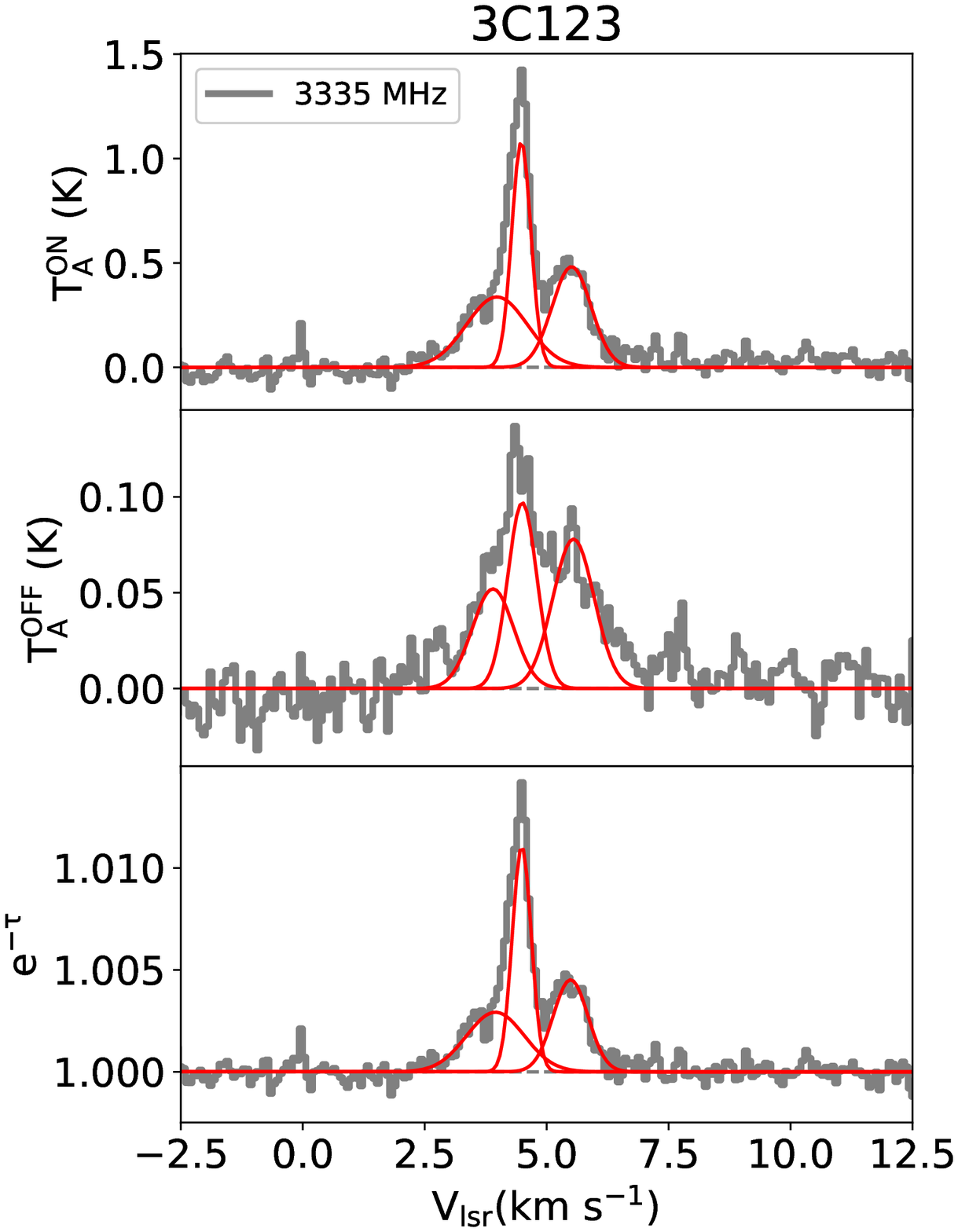}{0.32\textwidth}{(b)}
              \fig{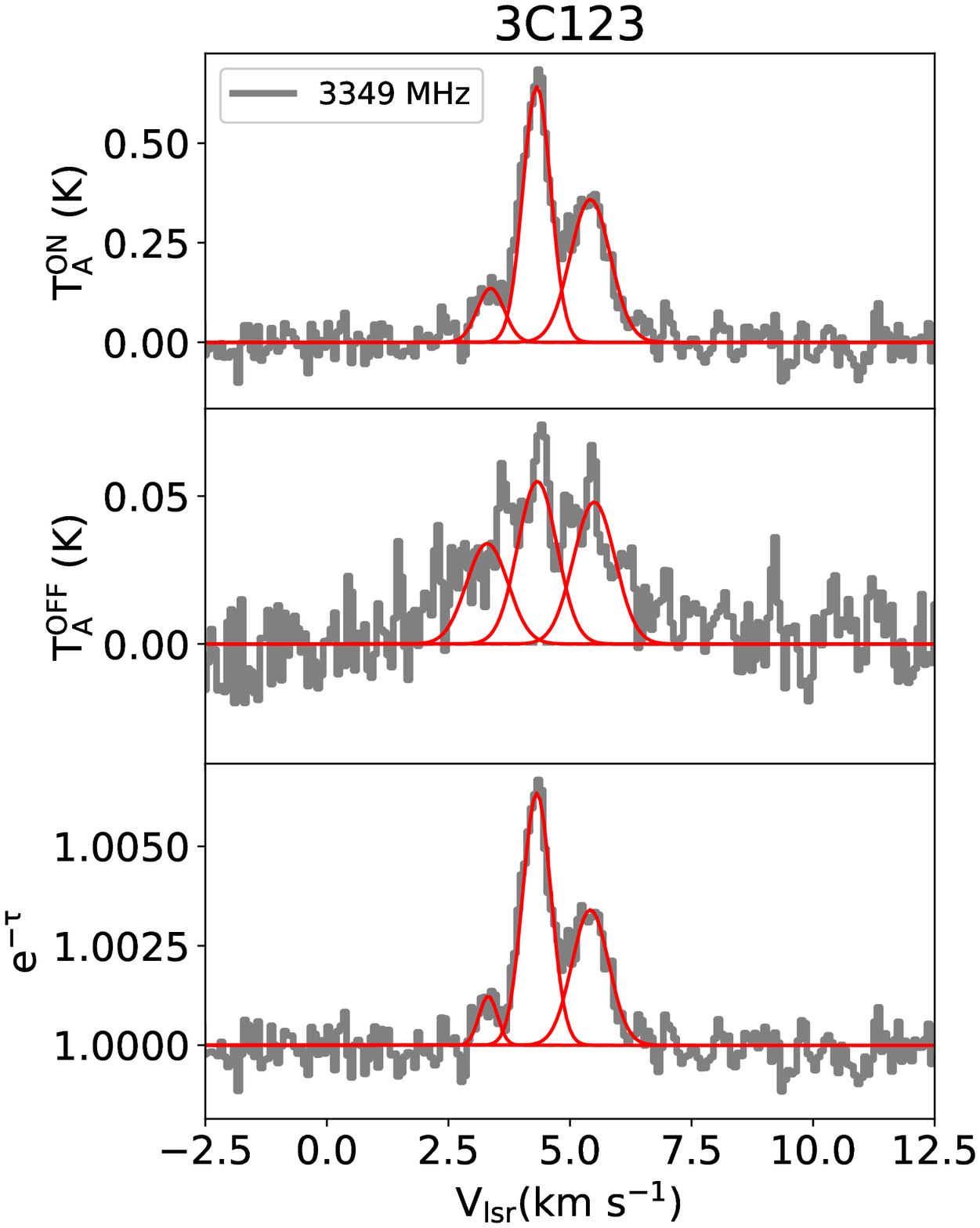}{0.32\textwidth}{(c)}
              }
\caption{Spectra and Gaussian decomposition results of 3C123 for 3264  (a), 3335 (b)  and 3349 MHz (c) line.  The top and middle panel represents the ON and OFF profile (antenna temperature) after removing the continuum level. The bottom panel represents the profile of optical depth, $e^{-\tau}$.}
\label{fig:3C123_profile} 
\end{figure*}

\begin{figure*}
\gridline{\fig{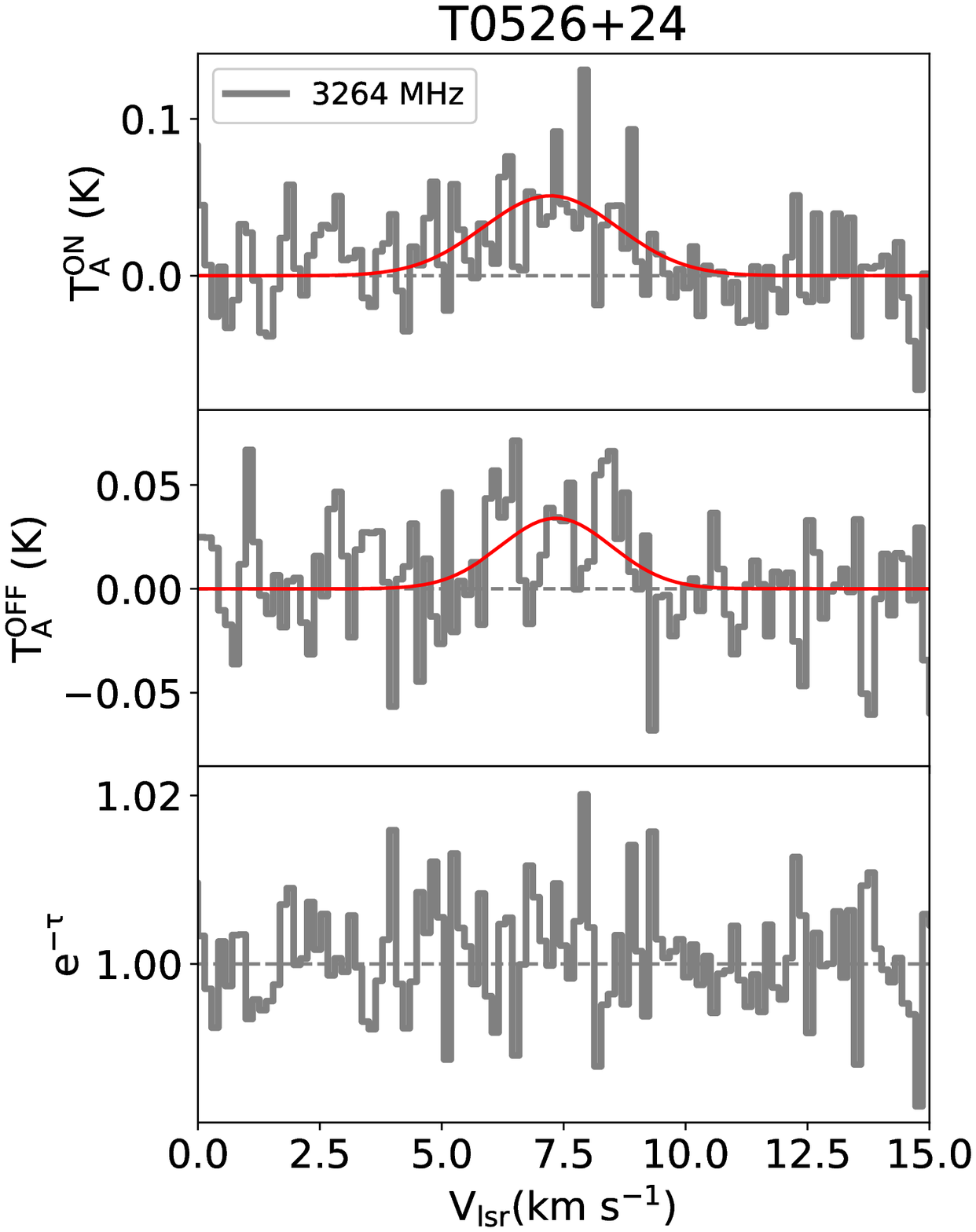}{0.32\textwidth}{}
              \fig{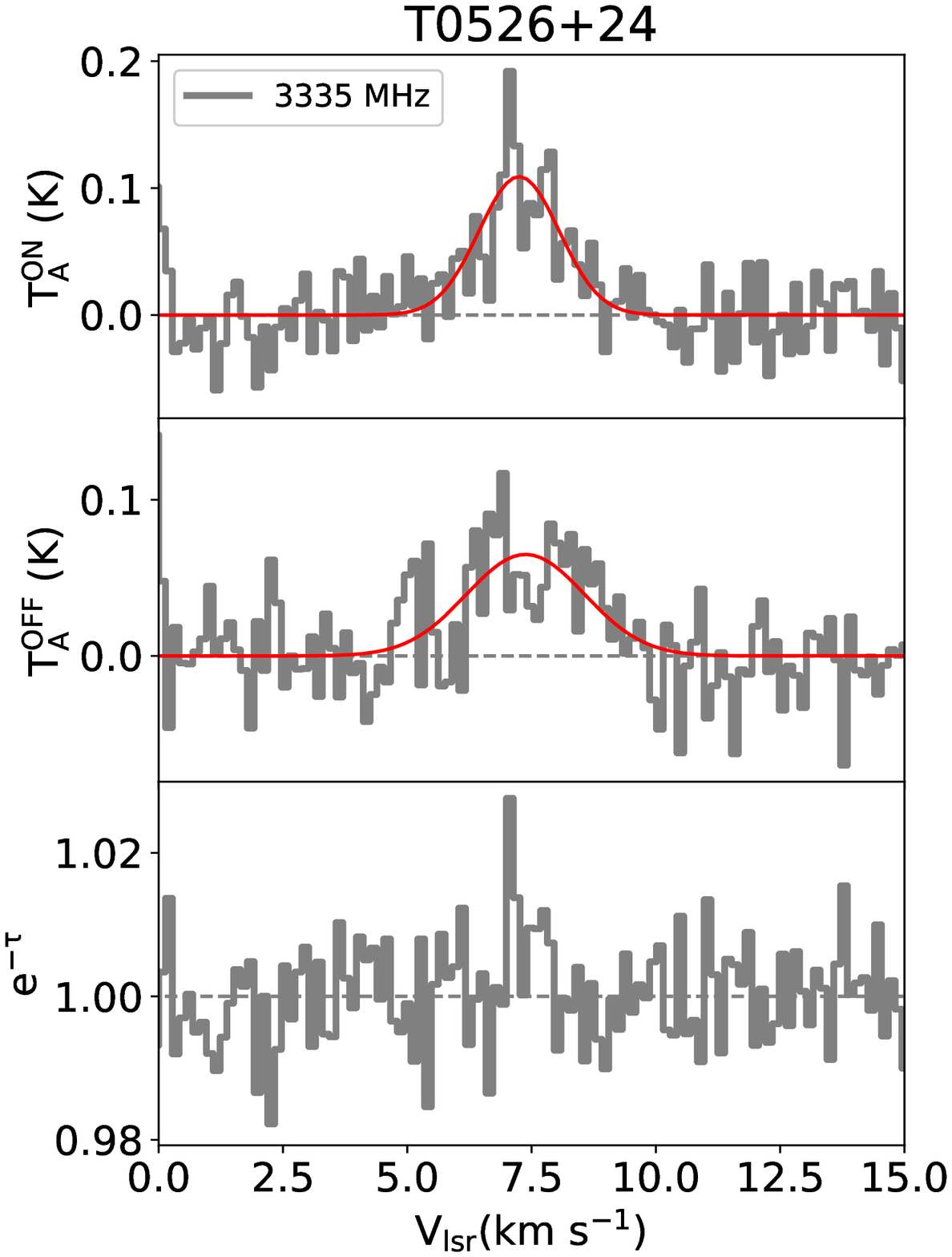}{0.32\textwidth}{}
              \fig{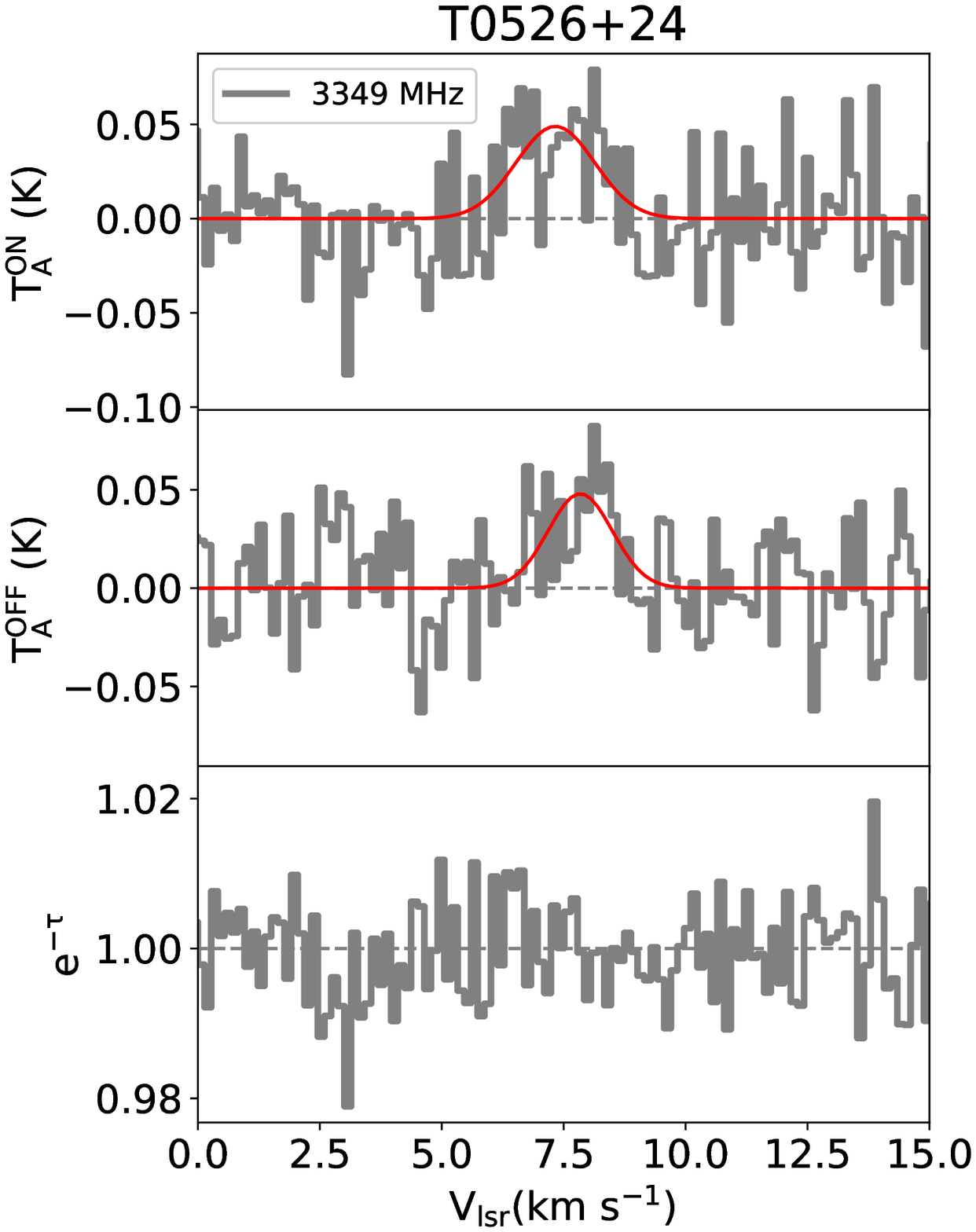}{0.32\textwidth}{}
              }
\gridline{\fig{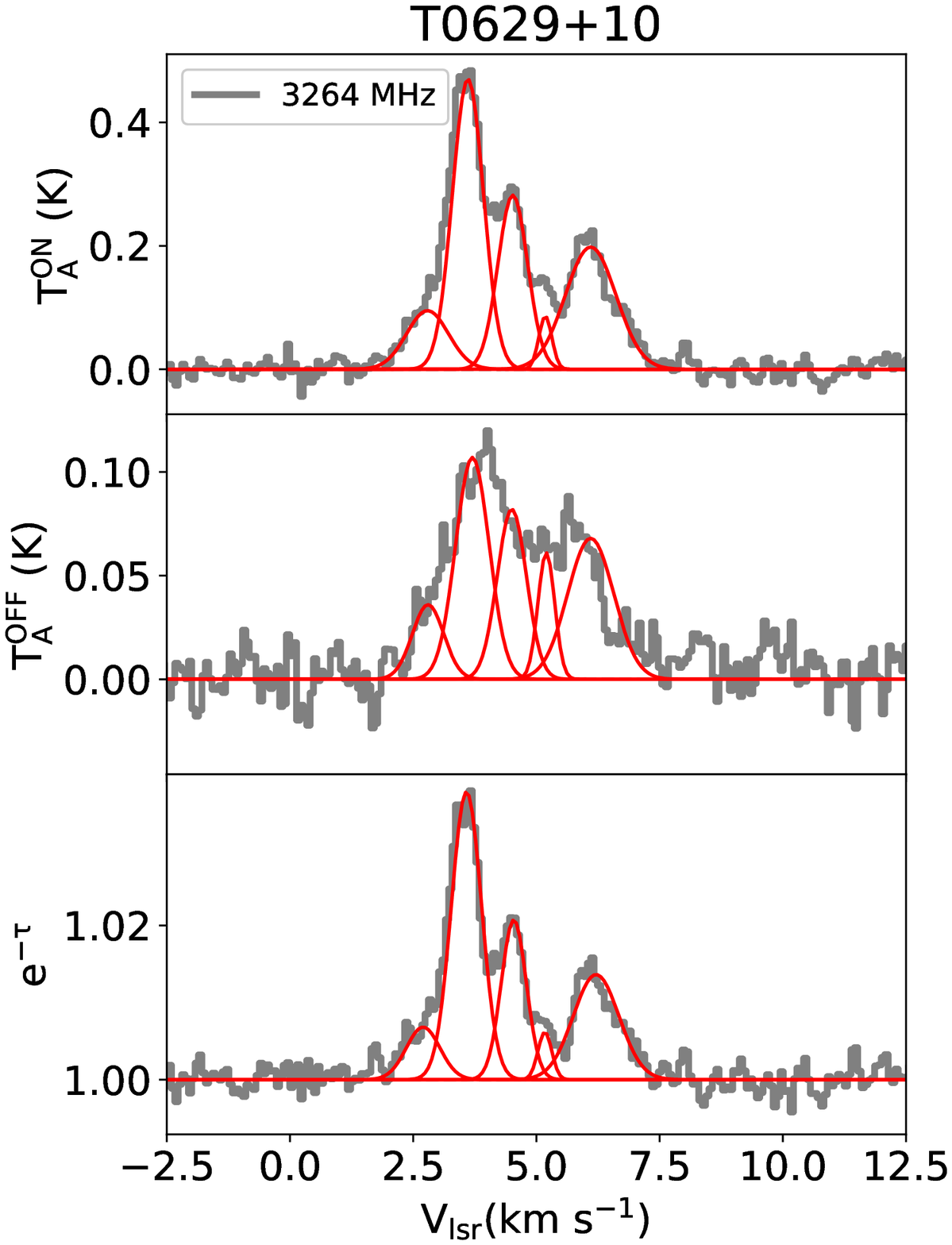}{0.32\textwidth}{}
              \fig{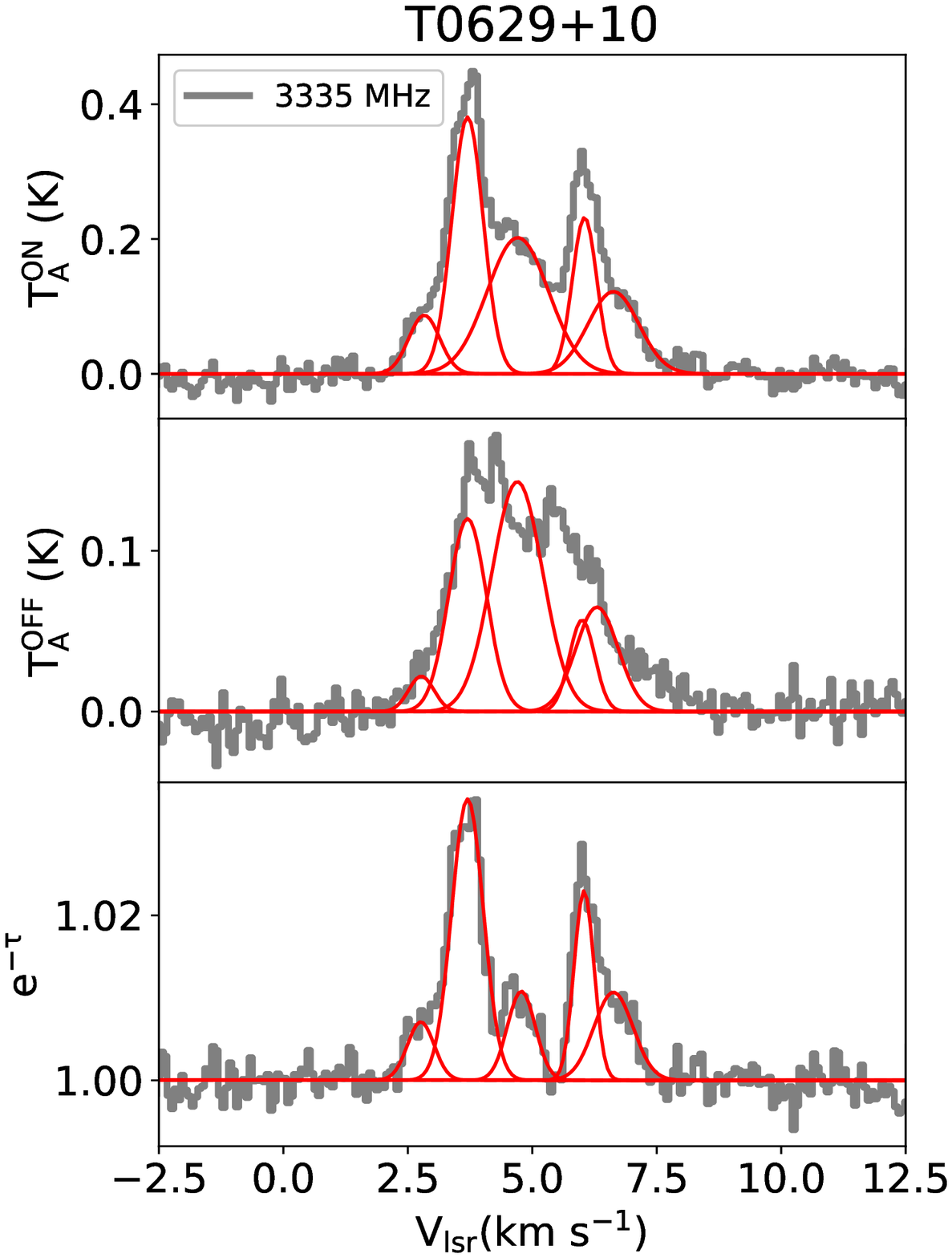}{0.32\textwidth}{}
              \fig{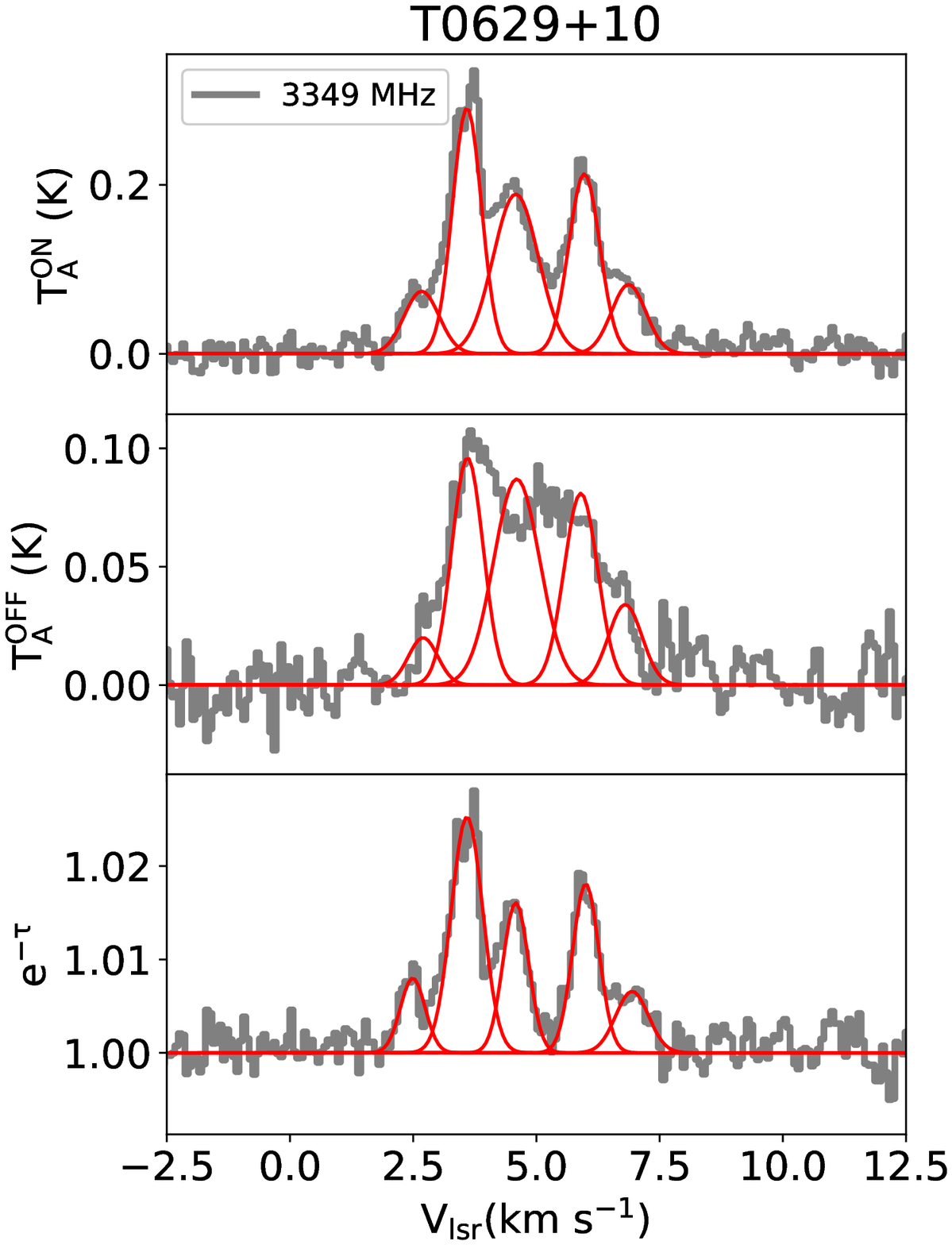}{0.32\textwidth}{}
              }
\caption{Spectra and Gaussian decomposition components of 3C131, 3C133, 3C154, T0526+24 and T0629+10. The description of each panel is same with that in the Figure \ref{fig:3C123_profile}.}
\label{fig:other_profiles}  
\end{figure*}

The Gaussian fitting results of three CH transitions toward 3C123 are shown in Figure \ref{fig:3C123_profile} as an example.  The fitting results of the rest sources can be found  in Figure \ref{fig:other_profiles}. The derived physical parameters (line central velocity, line width, optical depth, excitation temperature, and column density) are presented in Table \ref{table:chfitresult} and Table \ref{table:chfitresult1}.  A total of 12, 13, and 15 Gaussian components of 3264, 3335, and 3349 MHz lines are identified, respectively.

\section{Physical Properties of CH}
\label{sec:CH_properties}

The physical parameters were calculated based on the fitting results of Gaussian components. They are described as follows. 

\subsection{Excitation Temperature and Optical Depth}
\label{subsec:tex_distribution}

The ON/OFF observations provide a direct measurement of excitation temperature $T_{ex}$ and optical depth $\tau$ in the case of 1 and 2. For  case 3, lower limits are derived due to the absence of CH detection in the OFF  position.   

\begin{figure*}
\gridline{\fig{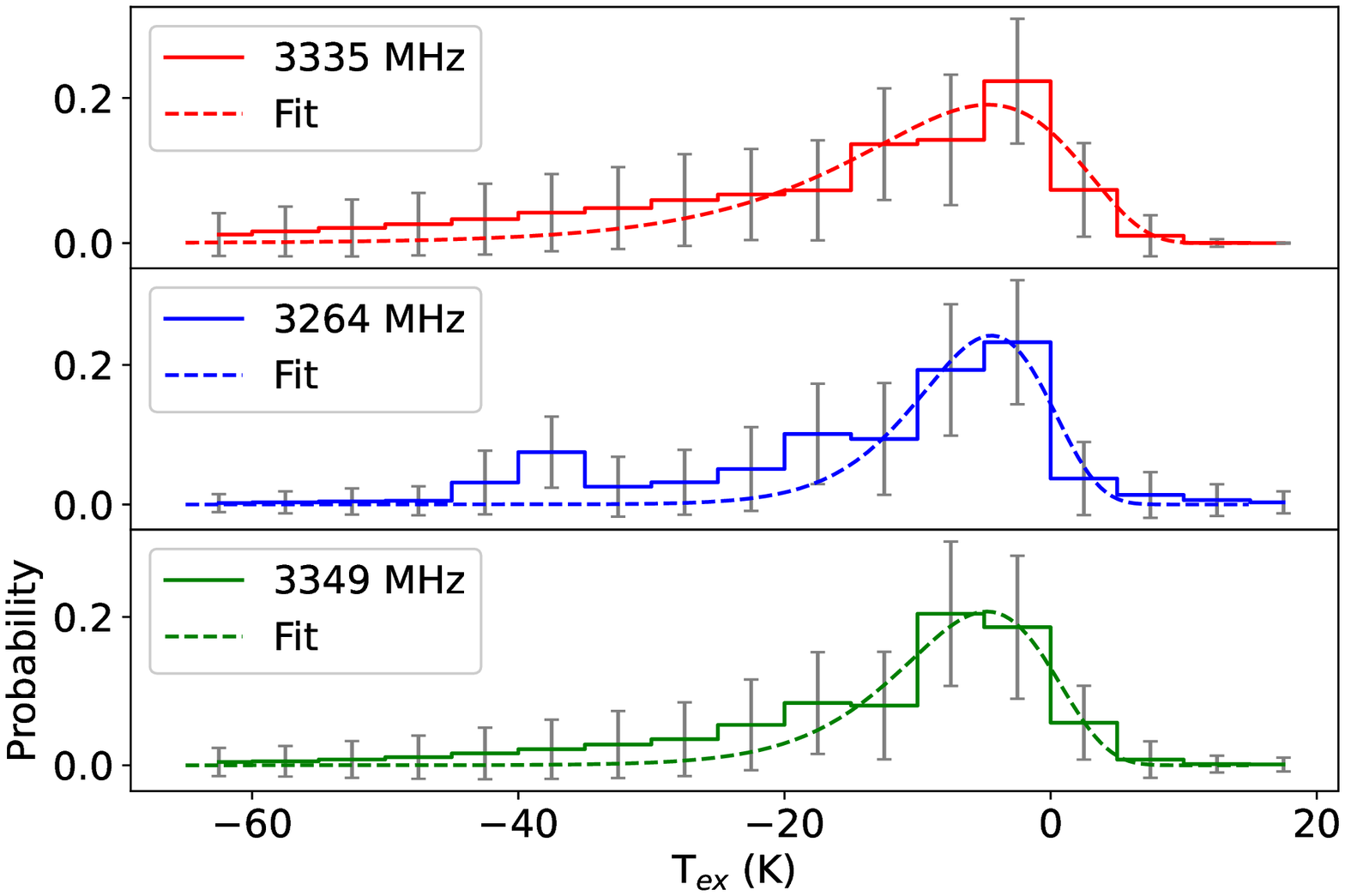}{0.49\textwidth}{(a)}
              \fig{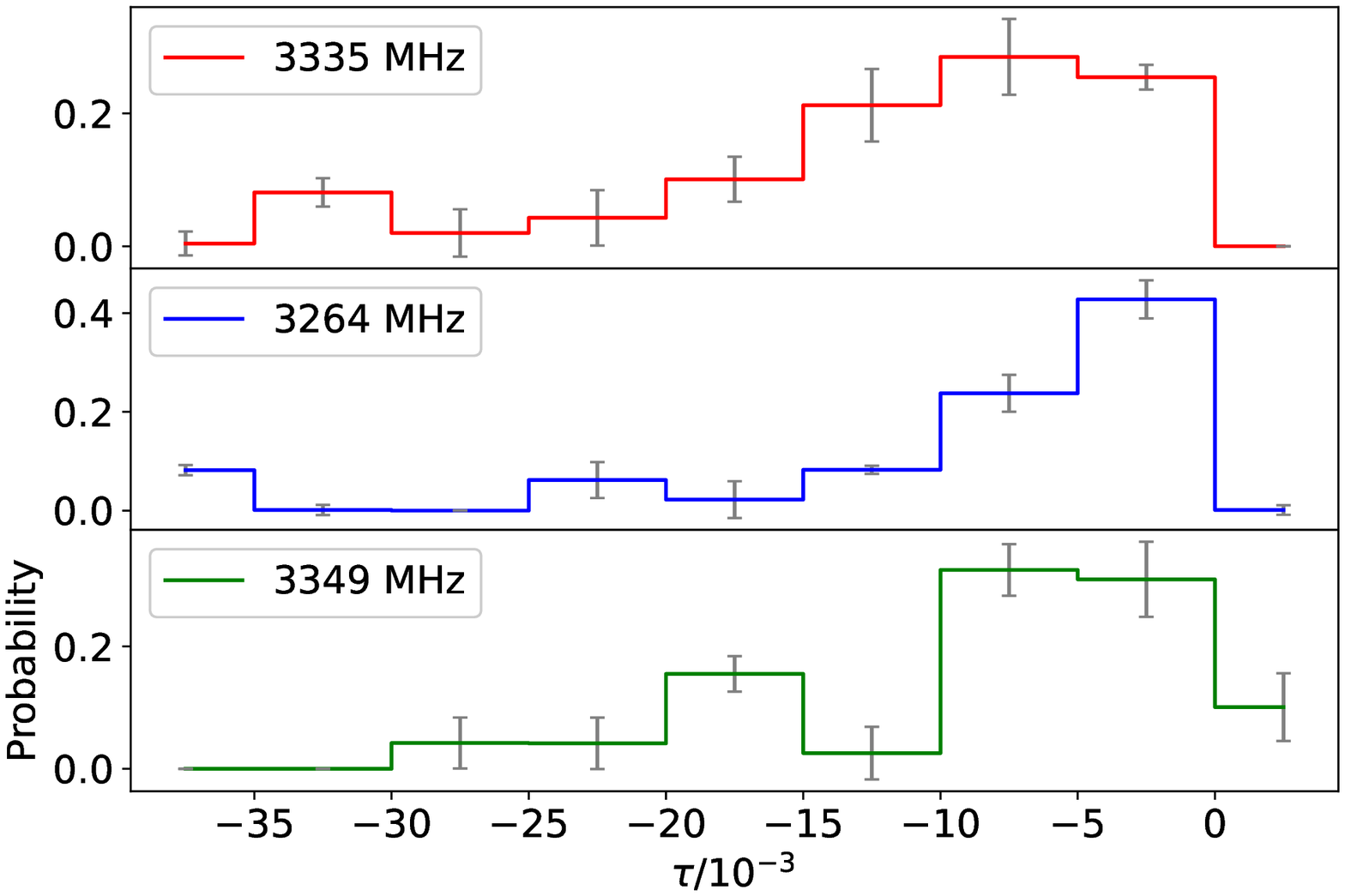}{0.49\textwidth}{(b)}}
\caption{ Histogram of  excitation temperature (a) and optical depth (b)  of CH transitions. The dashed lines in figure (a) represent fitting results of excitation temperature distribution.The two large T$\rm_{ex}$ values (300 and -754 K) of 3349 line are not included in the figure (a). }
\label{fig:ch_hist}
\end{figure*}

Due to the limitation of a  small sample number, the Monte Carlo method is  introduced to estimate the probability distribution of physical parameters.  
As shown in the histogram in Figure \ref{fig:ch_hist}(a), the excitation temperature peaks of CH 3335\,MHz and 3264\,MHz lines are in the range of [$-5$,0] K, which is consistent with that found by combining the optical and radio CH observations \citep{2020MNRAS.495..510D}. For CH 3349\,MHz line, the excitation temperature peak is  in  more negative values with range of [$-10$, $-5$] K. Our results confirm that the excitation of CH 9\,cm transition is always strongly inverted (negative $T\rm_{ex}$ value). The distributions of excitation temperature $T\rm_{ex}$ with range of (-62, 15) K are fitted with a log-normal function, which follows

\begin{equation}
    f(x) = \frac{a}{(15-T_{ex})\sigma}e^{-\frac{(ln(15-T_{ex})-\mu)^2}{2\sigma^2}},
\end{equation}
where a, $\mu$ and $\sigma$ are three free parameters. The fitting results of a, $\mu$ and $\sigma$ for 3335, 3264 and 3349 MHz lines are (1.7, 3.2, 0.4), (1.2, 3.0, 0.3) and (1.2, 3.1, 0.3) respectively.

As shown in Figure \ref{fig:ch_hist}(b), the CH lines are  optically thin in our sample (with maximum  $|\tau|$ value of 0.034). The optical depth of 3264\,MHz, 3335\,MHz, and 3349\,MHz are in the range of  (-36.7$\sim$-2.3)$\times$10$^{-3}$, (-33.5$\sim$-2.9)$\times$10$^{-3}$, and (-25$\sim$0.2)$\times$10$^{-3}$, respectively.

The comparison with previous CH 3335 MHz observations of 3C123 and 3C133 is shown as follows.

3C123: As a bright calibrator,  3C123 has been observed under ON/OFF mode for many times in the last 45 years.  With the Onsala Space Observatory 25.6m telescope,  $T\rm_{ex}$  was estimated as $-10$ K in  \citet{1976ApJS...31..333R}   and $-9$ K in \citet{1977ApJS...35..263H}. \citet{2002A&A...391..693L} determined a consistent $T\rm_{ex}$ value of  $-10.7\pm$ 3.2 K with the  NRAO 43m telescope. Due to sensitivity limitation of these observations, only one component centering at $\sim$4 \kms\ was detected.  \citet{1979A&A....73..253G} identified two velocity components (centering at 4.5 and 5.5 \kms)  with the Effelsberg 100 m telescope, resulting in a $T\rm_{ex}$ of $-60 \pm 30$ K of the 3335 MHz main line. 

With improved sensitivity and spectral resolution of the Arecibo telescope, we have identified 3 velocity components (centering at 4.0, 4.5 and 5.5 \kms).  The corresponding $T\rm_{ex}$ values are $-54.5 \pm 12.1$, $-10.5 \pm 2.9$ and $-33.2 \pm 6.8$  K, respectively.  The first two values are consistent with previous studies. These results suggest  that the beam size and sensitivity are critical for determining the CH excitation temperature. 

3C133: It was observed by \citet{2002A&A...391..693L} but there is no clear implication of $T\rm_{ex}$ value toward this source.  In this work, we derived $T\rm_{ex}$ value of  $-13.3 \pm$ 2.4 K toward the $7.7\pm 0.0$ \kms component.  Compared to the 3335\,MHz main line,  the OFF spectrum of 3349\,MHz line shows an extra component with central velocity of $8.5 \pm 0.1$ \kms. The origin of this extra velocity component is difficult to understand now.  Further interferometric observations with high spatial resolution, e.g., VLA are needed. 

\subsection{CH  Column Density }
\label{subsec:ch_columnden}

The transition energy between the lowest two excitation states ( $J$=3/2 and $J$=1/2) is  25.5 K.  While the critical density is 10$^6$ \cc \citep{2012A&A...546A.103S},  meaning that almost all CH molecules are located in the J=1/2 $\Lambda$-type doubling levels in our sightlines.  With the assumption of local thermal equilibrium (LTE), the CH populations at different $F$ levels  follow the statistical population distribution since the excitation temperatures of 3264, 3335, and 3349\,MHz lines are same. The total CH column density under LTE condition can be derived  by  \citep{1979A&A....73..253G}

\begin{equation}
\rm N(CH)^{LTE} = 2.82\times 10^{14} cm^{-2}  \delta T_{ex}   \Delta V \tau  ,
\label{eq:NCH_LTE}
\end{equation}
where  $\Delta V$ is the FWHM of CH line respectively.  The statistical  ratio $\delta$ is 1 for the 3335 MHz line and  2 for the 3264 and 3349 MHz line \citep{1976ApJS...31..333R}.  

As presented  in section \ref{subsec:tex_distribution}, excitation temperatures of 3 transitions are totally different even when uncertainties are considered, indicating the invalidity of LTE assumption. In this case, the total column density should  be calculated through \citep{2011A&A...531A.121S}

\begin{equation}
\rm N(CH) = N_{u,1} (1 + e^{\frac{T_{11}}{T_{ex,11}}} + \frac{1}{3}e^{\frac{T_{10}}{T_{ex,10}}}  + \frac{1}{3}e^{\frac{T_{11}}{T_{ex,11}}-\frac{T_{01}}{T_{ex,01}}}),
 \label{eq:NCH_all}
 \end{equation}
 where $N_{u,1}$ is the upper $F$=1 column density of the J=1/2 ground state and is calculated from the 3335 MHz data, $N_{u,1}$= $N$(CH)$^{\rm LTE}$/2. $T_{ex,11}$, $T_{ex,10}$ and $T_{ex,01}$ are the excitation temperature of 3335, 3349 and 3264 MHz line respectively.
 
\begin{figure}
\begin{center}  
  \includegraphics[width=0.4\textwidth]{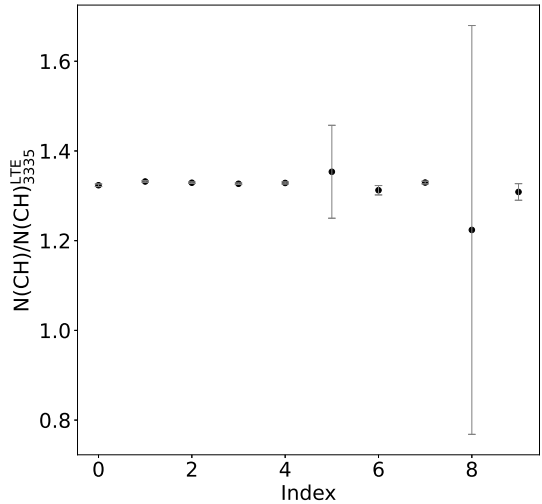}
\caption{Ratio between N(CH) and N(CH)$\rm^{LTE}_{3335}$ for 10 CH components. The ninth point with large error corresponds to the T0629+10 component centering at 6.1 \kms, which has $T\rm_{ex}^{3335}$=$-0.9\pm 4.1$ K.}
\label{fig:NCH_ratio} 
\end{center}
\end{figure}

With the direct measurement of both main line and satellite lines of CH, we are able to compare the estimating results from both Equation \ref{eq:NCH_LTE} and \ref{eq:NCH_all}.   Only 10 velocity components have optical depth and excitation temperature measurements of three CH lines. As shown in Fig. \ref{fig:NCH_ratio}, the ratios between N(CH) from Equation  \ref{eq:NCH_all} and N(CH)$\rm^{LTE}_{3335}$ based on 3335 MHz data and Equation \ref{eq:NCH_LTE} are almost constant, with a value of $1.32\pm 0.03$.  This implies that the previous studies of calculating N(CH) based on 3335 MHz data alone by adopting LTE assumption would underestimate  N(CH) by 32\%.  

 The column density of all the 12 CH 3335\,MHz components are shown in Table \ref{table:chfitresult}.  
The N(CH) for the two velocity components 3C123 (4.0 \kms) and 3C154 ($-2.2$ \kms) are estimated from N(CH)$\rm^{LTE}_{3335}$ by multiplying a factor of $1.32\pm 0.03$. N(CH) ranges over two orders of magnitude in our sample, from  $1.3\times 10^{12}$ to $1.4\times 10^{14}$ \cm2.

\subsection{Implication for N(CH) from CH 3335 MHz Emission}
\label{subsec:ch_imply}

Previous CH 3335 MHz observations (see section \ref{subsec:XCH} for details) always adopted a fixed temperature $T\rm_{ex}=-10 $ or  $-60$ K for the calculation of N(CH) through $N(CH) \propto (T_{ex}/(T_{ex}- T_{bg})) W(\rm CH)$, in which W(CH) is the integrated intensity of CH 3335\,MHz emission. 

\begin{figure}
\begin{center}  
  \includegraphics[width=0.4\textwidth]{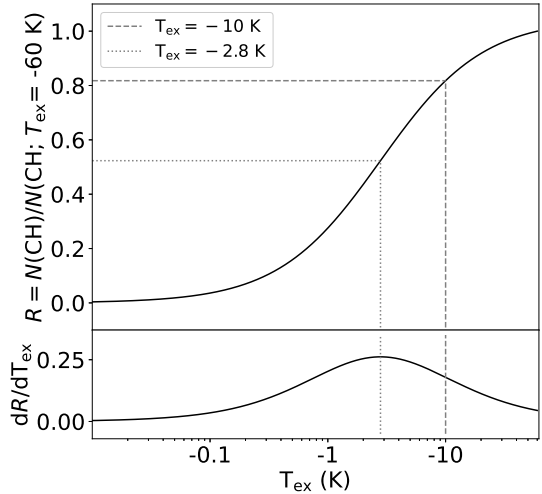}
\caption{ The value of $R=N$(CH)/$N$(CH; $T\rm_{ex}$= -60 K) (upper panel) and its first-order derivative (lower panel) as a function of excitation temperature. The background temperature is adopted as T$_{bg}=2.8$ K in our calculation. When $T\rm_{ex}=-2.8$ K (grey dotted line),  the $R$ value is 0.52 and the d$R$/d$T\rm_{ex}$ value reaches its maximum. The corresponding $R$ value is 0.82 when $T\rm_{ex} =  -10.0$ K (grey dashed line).  }
\label{fig:ratio} 
\end{center}
\end{figure}

We define  $R$ as the ratio between $N$(CH)  and  $N$(CH; $T\rm_{ex}=-60$ K) calculated by adopting $T\rm_{ex}=-60 $ K, namely, $R=N$(CH)/$N$(CH; $T\rm_{ex}= -60$ K). As shown in Figure \ref{fig:ratio}, the value of $R$ decrease monotonously as a function of $T\rm_{ex}$.  The adoption of $T\rm_{ex}=-60 $ K and $T\rm_{ex}=-10 $ K would overestimate $N(\rm CH)$  by a factor of 2 and 1.6 respectively compared to  that of $T\rm_{ex}=-2.8$ K. A larger negative $T\rm_{ex}$ value ($-2.8$ K $<T\rm_{ex}< 0$ K) leads larger deviation. The overestimation factor is 9.9 and 8.2 for $T\rm_{ex}=-0.3$ K, which is found in \citet{2021A&A...650A.133J}.

\section{Relationship between CH and OH}
\label{subsec:ch_oh}

The two simple hydrides, CH and OH are two of the most popular molecular tracers in diffuse and translucent clouds. The abundance and relationship between the two species have been investigated by a lot of studies in both radio and visible wavelengths. 

\citet{1976ApJS...31..333R} obtained CH 9 cm and OH 18 cm absorption and emission spectra toward strong continuum sources and derived N(CH)/N(OH) =0.06 for compact molecular clouds and  N(CH)/N(OH) = 0.4 for the dilute \hi\ clouds.  \citet{2002A&A...391..693L}  summarized the CH observations of CH B-X 3890 \r{A} and C-X 3150 \r{A}  \citep{1994ApJ...424..772F, 1995ApJS...99..107C} and the corresponding electronic OH  A-X and D-X  observations \citep{1996MNRAS.279L..37R,1996ApJ...465L..57F}  along 4 sightlines (Table 1 of \citet{2002A&A...391..693L} ), in which they found the N(CH)/N(OH) ratio is  0.33$\pm$ 0.10.  \citet{2009A&A...499..783W}  and \citet{2010MNRAS.402.1991W}  obtained  CH and OH column densities through  CH B-X data at 3886 and 3890 \r{A} and OH A-X data at 3078 \r{A} and 3083 \r{A} toward a total of 20 translucent sightlines (\citet{2002A&A...391..693L} adopted some of their data), in which the OH data toward 4 sightlines were obtained from \citet{1996MNRAS.279L..37R}, \citet{1996ApJ...465L..57F} and \citet{2005A&A...429..509B}. They found the column density of CH and OH have positive correlation,  N(CH)= (0.35$\pm$ 0.015)N(OH) + 0.40$\pm$ 0.17 (in 10$^{13}$ \cm2), in which the column density are calculated for the line-of-sight, not for different velocity components. The average abundance ratio of N(CH)/N(OH) is $0.40\pm 0.06$.

\citet{2018ApJS..235....1L} presented OH column density N(OH) of decomposed velocity components toward the sightlines in this work.  By comparing the  same velocity components, we found that most CH components satisfy N(CH)$<$ N(OH) except for two, T0526+24 centering at  7.2 \kms and T0629+10 centering at  4.7 \kms. The N(CH)/N(OH) ratios of the two components are $6.4\pm 5.2$ and $10.8\pm 6.5$, implying an overabundance of CH in these two clouds. This phenomenon of enhanced CH abundance was also detected toward the source HD 34078 in  \citet{2009A&A...499..783W} and will be discussed in the section \ref{overabunt_ch}. When these two velocity components are excluded, the N(CH)/N(OH) ratio of the  9 clouds ranges from 0.06 to 0.83 with an average value of $0.44 \pm 0.28$. 

\begin{figure}
\begin{center}  
  \includegraphics[width=0.45\textwidth]{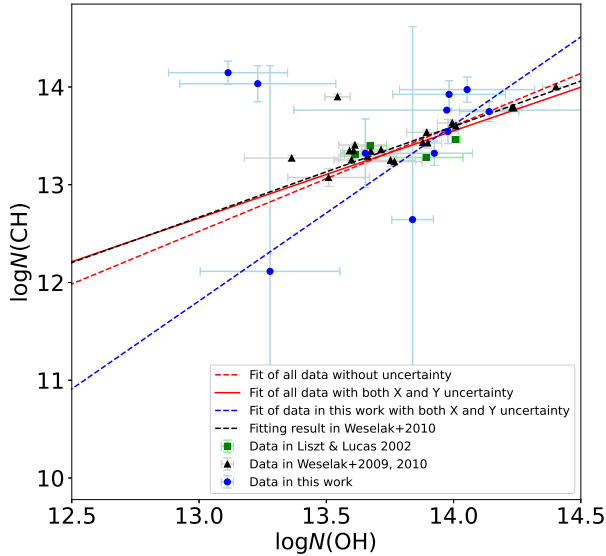}
\caption{ Relationship between $N$(CH) and $N$(OH).  Compared to radio data in this paper, the data of  Weselak et al. (2009, 2010) were obtained from visible  spectra.  The dashed black line represents the fitting result given in Weselak et al.\ (2010), in which the  source HD 34078 (upper left in the figure) with enhanced CH abundance was not included.  The blue dashed line represents fitting result of data in this work, in which the two clouds with overabundant CH (T0526+24 centering at  7.2 \kms and T0629+10 centering at  4.7 \kms, upper left in the figure) are excluded.  The fitting result (red solid line) of all data does not include these 3 points.}
\label{fig:ohch} 
\end{center}
\end{figure}

As shown in the Figure \ref{fig:ohch}, the relationship  between N(CH) and N(OH) shows a tight correlation for the 9 clouds. Due to the presence of uncertainties in both N(CH) and N(OH), we introduced the Orthogonal Distance Regression (ODR) method instead of Ordinary Least Square fitting to fit the relationship. 
The column density of CH and OH could be linearly fitted with the following formula, log$N$(CH) = (1.80$\pm$ 0.49) log$N$(OH) $-11.59 \pm 6.87$.  After excluding 3  data points whose uncertainty value is larger than their data value, the ODR fit for the rest 6 data points is log$N$(CH) = (1.76$\pm$ 0.64) log$N$(OH) $-10.99 \pm 8.92$.  This means that the 3  data points with large uncertainties contribute little to the trend of CH-OH relationship. 

We combine the results with explicit measurement from \citet{2002A&A...391..693L},  \citet{2009A&A...499..783W, 2010MNRAS.402.1991W},  and this work to construct a full sample of 32 points, in which the abnormal 3 points are excluded.  The N(CH)/N(OH) ratio of this sample is $0.43\pm 0.18$.   After adopting ODR method, the relationship between CH and OH could be fitted with  log$N$(CH) = (0.89$\pm$ 0.09) log$N$(OH) $+ 1.06 \pm 1.21$. For comparison, the fit without uncertainty leads to log$N$(CH) = (1.08$\pm$ 0.17) log$N$(OH) $ -1.48 \pm 2.42$.  

\subsection{Overabundance of CH}
\label{overabunt_ch}

As shown in section \ref{subsec:ch_oh}, the overabundance of CH exists.  Due to the higher reaction rate of CH formation at higher temperature, increasing the kinematic temperature (e.g., $\sim$ 300 K) in PDR model with equilibrium chemistry could greatly increase N(CH). But this process would also accelerate the reaction rate of other species (e.g., OH, CO) \citep{1988ApJ...334..771V}. 

Overabundant CH$^+$ has been observed by multiple observations  \citep[e.g.,][]{2008ApJ...687.1075S}. The overabundance of $\rm CH^+$ is able to accelerate  CH formation through the reaction $\rm CH^+ + e \rightarrow CH$ or even faster reactions  CH$^+$ + \h2\ $\rightarrow$ CH$_2^+$ + H  and CH$_2^+$ + e $\rightarrow$ CH + H.  However, the formation of CH$^+$ in the diffuse ISM is difficult since the reaction, C$^+$ + \h2\ $\rightarrow$ CH$^+$ + H, is endothermic with a rate of $1\times 10^{-10}$exp(-4640/T$\rm_{kin}$) \cc s$^{-1}$ \citep{1996MNRAS.279L..41F}.   

The nonequilibrium chemistry, single-fluid shock models  \citep{1978ApJ...222L.141E} is able to produce observed CH$^+$ abundance but results in overabundant OH. The two-fluid magnetohydrodynamic (MHD) shock (C-shock) model  \citep{   1986ApJ...306..655D, 1986ApJ...310..392D, 1986ApJ...310..408D}  was introduced to explain the CH$^+$ abundance problem.  The main problem of this model is that the predicted velocity shift  \citep[5 \kms,][]{1991ApJ...375..642H}  between CH$^+$ and CH absorption profile was not observed by high resolution observations  \citep{1993A&A...269..477G, 1995ApJS...99..107C, 1995MNRAS.277..458C, 1997A&A...320..929G}.  \citet{1998MNRAS.297.1182F}  claimed that the velocity shift could be less than 2 \kms from the  C-shock model. With $n\rm_H=50$ \cc, fractional ionization $n\rm_e$/$n\rm_H$ = $2.3\times 10^{-4}$, magnetic field $B=7.1$ $\mu$G and UV intensity $\chi=3$,  the N(CH)/N(OH) value reaches  $\sim 14$ \citep{1986ApJ...310..392D}, which  could explain the observed values in this work.  Although no such shift was observed between the central velocities of OH and CH, one cannot rule out the shock origin of the CH and CH$^+$ over-abundance, considering the geometrical reliance of shock models. \citet{2016ApJ...833...90X} has found the C-shock model to be a probable explanation for over-abundant CH in the transition zone across the Taurus molecular cloud edge. Clearly, more systematic surveys and analyses are required to ascertain the role of shocks in CH excitation. The current and upcoming SKA precursors (ASKAP and MeerKAT in particular) and SKA-1 poise to fulfill such requirements. We particularly note the successor of Arecibo, the Five-hundred-meter Aperture Spherical radio Telescope  \citep[FAST:][]{2018IMMag..19..112L,2019RAA....19...16L}. We are expanding the Millennium survey with FAST. The upper end of the FAST frequency coverage is currently at 1.5 GHz, but expected to be extended to 3.3 GHz in about a year. The improved sky-coverage and the sensitivity of FAST will facilitate substantial increase of HI, OH, and CH absorption measurements, over the Millennium survey. 

\section{CH  vs. \h2}
\label{subsec:XCH}

CO low-$J$ rotational transitions fail to trace diffuse molecular gas and the so called ``dark gas" \citep[e.g.][]{2005Sci...307.1292G, 2011A&A...536A..19P} because of the low excitation and low abundance.
 \citet{1973ApL....15...79B} and \citet{1975ApJ...199..633B} first investigated the chemical models leading to CH formation through reactions with \h2. These models imply a close correlation between CH and \h2, which was found to be consistent with observations \citep{1982ApJ...257..125F}.  Many efforts have been done to measure the relationship between column density of CH and \h2 in various environments. We summarized the results derived by a series of radio, visible, millimeter and sub-millimeter  observations toward clouds with different environments:  

\begin{enumerate} 
\item  By assuming $|T\rm_{ex}| =-15$ K,  \citet{1981A&A....97..317S}  found $N\rm(CH)\approx 6\times 10^{13} A_B$ through 3335 MHz emission toward L1642. 

\item  With the assumption of $T\rm_{ex}=-60$ K,  \citet{1986A&A...160..157M}  obtained the abundance N(CH)/N(\h2)= $4.0\times 10^{-8}$ and correlation log$N$(CH)= (1.02$\pm$ 0.04) log$N$(H$_2$) $- 7.82$ based on 3335 MHz observations of low extinction clouds L134, L1780 and L1590. This relationship is valid for $10^{19.5}<N(H_2)<10^{21.5}$ \cm2 and  breaks down for dense gas regions with $N(H_2)\gtrapprox 3\times 10^{22}$ \cm2.

\item  Based on 3264, 3335, and 3349 MHz emission data, \citet{2011A&A...531A.121S}  found  that CH abundance in TMC1 ranges from $1.0\times 10^{-8}$ to $\sim 2.2\times 10^{-8}$.  The value $T_{ex}^{3335}=-10 $ K was chosen to calculate N(CH). 

\item  \citet{2016ApJ...833...90X}  investigated CH emission across the Taurus boundary and found N(CH) as a function of visual extinction A$\rm_V$, N(CH)= 2.6$\times 10^{13}$exp[($-(\frac{A_V-2.0}{1.0})^2)$]+ $1.6\times 10^{13}$ \cm2. The assumption of $T_{ex}^{3335}=-60 $ K was adopted when calculating N(CH). 

\item  Visible transitions of CH and  Lyman $B-X$ transitions of \h2\  provide direction measurement of N(CH) and N(\h2), respectively.  \citet{2002A&A...391..693L} collected previous direct N(CH) and N(\h2) measurements \citep{1995ApJS...99..107C, 1995MNRAS.277..458C, 1994ApJ...424..754A, 1993A&A...269..477G,  1994ApJ...424..772F,   1984A&A...130...62D, 1989ApJ...340..273V, 1985ApJ...288..604C,  1992A&A...265L...1J, 1993ApJS...88..433P, 2001ApJ...555..839R, 1977ApJ...216..291S}  and found averaged CH abundance X$\rm_{CH}$=N(CH)/N(H$_2$)= $4.3\pm 1.9 \times 10^{-8}$.  In addition, the CH and \h2\ relation follows log$N$(CH) = (1.00$\pm$ 0.06)log$N$(\h2) $-7.35\pm$ 1.31. 

With new visible  and far-UV measurements,  \citet{2008ApJ...687.1075S} found N(CH)/N(\h2)= $3.5^{2.1}_{-1.4}\times 10^{-8}$. The average relationship between N(CH) and N(\h2) for all data is  log$N$(CH)= ($0.97\pm 0.07$) log $N$(\h2) $-6.80\pm 1.50$. It is log$N$(CH)= ($0.92\pm 0.19$) log$N$(\h2) $- 5.87\pm 3.78$ when log$N$(\h2) $<20.4$ and log$N$(CH)= ($1.09\pm 0.19$) log$N$(\h2) $- 9.34\pm 3.90$ when log$N$(\h2) $>20.4$. 

\end{enumerate}

Traditionally, the proton column density was derived from E(B-V) value through N$\rm_H$= 5.8$\times 10^{21}$E(B-V) \citep{1978ApJ...224..132B}.  More and more evidences present a higher conversion factor. By comparing precise N(\hi) and E(B-V)  for pure atomic sightlines of Millennium survey,  \citet{2018ApJ...862...49N}  found the  conversion ,  N$\rm_H$= (9.4 $\pm$ 1.6)$\times 10^{21}$E(B-V),  which is  62\% higher compared to that in \citet{1978ApJ...224..132B}.  
With adoption of this conversion value, it is able to derive  total column density along sightline $N\rm_H^{LOS}$ based on E(B-V) measurement from  \citet{1998ApJ...500..525S}.  The column density of all decomposed \hi\ components   are present in  \citet{2003ApJ...586.1067H}.  In order to calculate total proton in the detected CH clouds, we subtracted N(\hi) contribution of  \hi\ components that do not cover the velocity range of CH emission.

\begin{figure*}
\begin{center}  
  \includegraphics[width=0.7\textwidth]{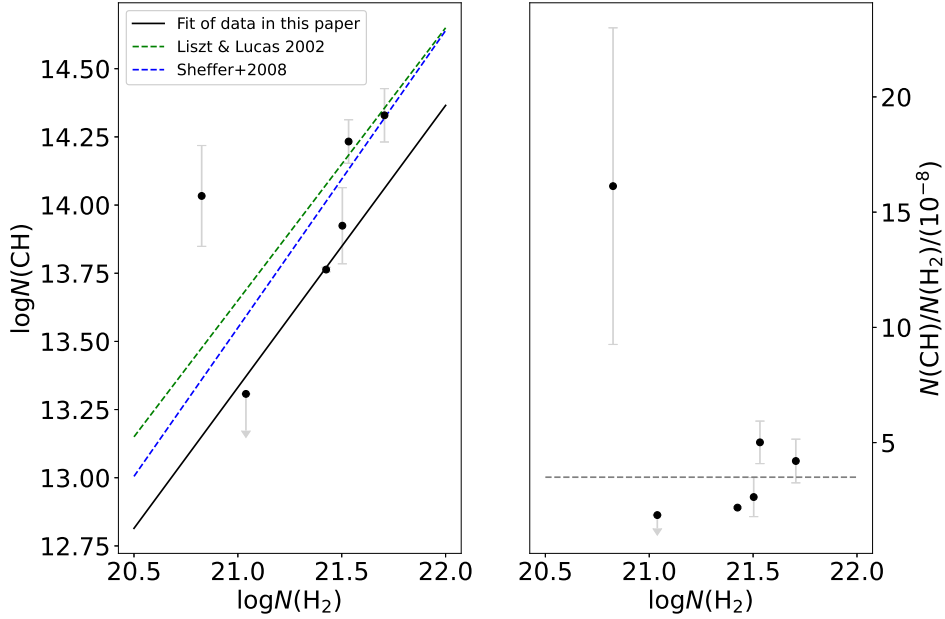}
\caption{ Relationship between $N$(CH) versus $N$(\h2) ($Left\ panel$) and N(CH)/N(\h2) versus $N$(\h2) ($Right\ panel$).  The solid line in the left panel represents linear fitting result, log$N$(CH)= (1.03$\pm$ 0.66) log$N$(\h2) $-$8.38 $\pm$ 14.16. The source T0526+24 with significant deviation of N(CH)/N(\h2)  was excluded during fitting.  The green and blue  dashed lines represent the fitted profile  in  Liszt \& Lucas (2002) (log$N$(CH) = (1.00$\pm$0.06) log$N$(\h2) $-7.35 \pm 1.31$) and  Sheffer et al.\ (2008) (log$N$(CH)= ($1.09\pm 0.19$) log$N$(\h2) $- 9.34\pm 3.90$). The dashed line in the right panel represents the value of 3.5$\times 10^{-8}$. The point with upper limit is 3C154. }
\label{fig:chh2} 
\end{center}
\end{figure*}

The column density of \h2 is calculated by $N$(\h2) = ($N\rm_H^{LOS}- N\rm^{tot}$(\hi))/2, in which N$\rm^{tot}$(\hi) is total \hi\ column density along sightline.  As shown in the left panel of Figure \ref{fig:chh2}, the relationship between N(CH) and N(\h2) follows log $N$(CH)= (1.03$\pm$ 0.66) log $N$(\h2) $-$8.38 $\pm$ 14.16.  Due to the limit of sample number, the slope of 1.03$\pm$ 0.66 has a relative large uncertainty but is consistent with that derived from visible and far-UV absorptions \citep{2002A&A...391..693L, 2008ApJ...687.1075S}. 
 
The right panel of  Figure  \ref{fig:chh2} shows the relationship between CH abundance X(CH)=N(CH)/N(\h2) and N(\h2). Only upper limit of X(CH) was derived toward the source 3C154. The mean value of X(CH)  is $6.0\pm 1.4 \times 10^{-8}$ toward other five sources (3C123, 3C131, 3C133, T0526+24, and T0629+10).  It becomes  $3.5\pm 0.4\times 10^{-8}$ when the source T0526+24 (X(CH)= $16.1\pm 6.9 \times 10^{-8}$) is excluded. Though the mean X(CH) value of $3.5\pm 0.4\times 10^{-8}$ is highly consistent with X(CH)= $3.5^{2.1}_{-1.4}\times 10^{-8}$ in \citet{2008ApJ...687.1075S}, more statistical samples are necessary. 

\section{Summary}
\label{sec:summary}

We have performed high-sensitivity CH observations toward 18 extra-galactic continuum sources with the Arecibo telescope.The excitation temperature and optical depth of CH are obtained directly through ON/OFF observations.Combining with the existing OH, \hi\, and extinction data, the main conclusions are as follows:

\begin{enumerate}

\item  Although spreading in a wide range between $-54.5$ K to $-0.5$ K,  23\% of the excitation temperatures of the CH 3335 MHz main line concentrate within [$-5, 0$] K according to our Monte Carlo analysis (Figure \ref{fig:ch_hist} (a)).  Compared to the adoption of $T\rm_{ex}= -3$ K, $N$(CH) derived with $T\rm_{ex}=-10 $ K (Rydbeck et al.\ 1976) and $T\rm_{ex}=-60 \pm 30$ K (Genzel et al.\ 1979) will be overestimated by a factor of 1.7 and 2.0, respectively. 

\item  The CH excitation temperatures can be described by a log-normal distribution, 
\begin{equation}
    f(x) = \frac{a}{(15-T_{ex})\sigma}e^{-\frac{(ln(15-T_{ex})-\mu)^2}{2\sigma^2}}.
\end{equation}
The fitting values of a, $\mu$ and $\sigma$ for 3335, 3264 and 3349 MHz lines are (1.7, 3.2, 0.4), (1.2, 3.0, 0.3) and (1.2, 3.1, 0.3) respectively.

\item  The optical depth of 3335 MHz transition ranges from $-33.5\times 10^{-3}$ to $-2.9\times 10^{-3}$ (Figure \ref{fig:ch_hist} (b)), indicating that the CH emission is optically thin.

\item  We derived accurate N(CH) value based on three transitions of CH. The  value N(CH) will be underestimated by a factor of $1.32\pm 0.03$ when only 3335 MHz data is adopted under the  LTE condition (Figure \ref{fig:ratio}). 

\item  In this study, log$N$(CH) correlates linearly with log$N$(OH) following log$N$(CH) = (1.80$\pm$ 0.49) log$N$(OH) $-11.59 \pm 6.87$ for 9 data points. The relationship becomes log$N$(CH) = (0.89$\pm$ 0.09) log$N$(OH) $+ 1.06 \pm 1.21$ for a full sample of 32  data points when previous results are included (Figure \ref{fig:ohch}). 

\item The average CH abundance relative to \h2\  is $3.5\pm 0.4 \times 10^{-8}$ when the source T0524+10 with large deviation is excluded. The relationship between $N$(CH) and $N$(\h2) is found to be log$N$(CH)= (1.03$\pm$ 0.66) log$N$(\h2) $-$8.38 $\pm$ 14.16, consistent with result from  visible observations (Figure \ref{fig:chh2}).  

\end{enumerate} 

\section*{Acknowledgments}
We are grateful to the anonymous referee for the constructive suggestions, which have greatly improved this paper.  We thank Chao-Wei Tsai for useful discussion.
This work is supported by the National Natural Science Foundation of China (Grant No. 11988101, 11803051,  11725313,  and 12033005), National Key R\&D Program of China (2017YFA0402600 and 2018YFE0202900),  CAS International Partnership Program (114A11KYSB20160008).  

Finally, we would like to express our personal gratitude to the great Arecibo telescope, which  bestowed upon us not only fantastic data, but also life-long inspiration.

\begin{figure*}
\gridline{\fig{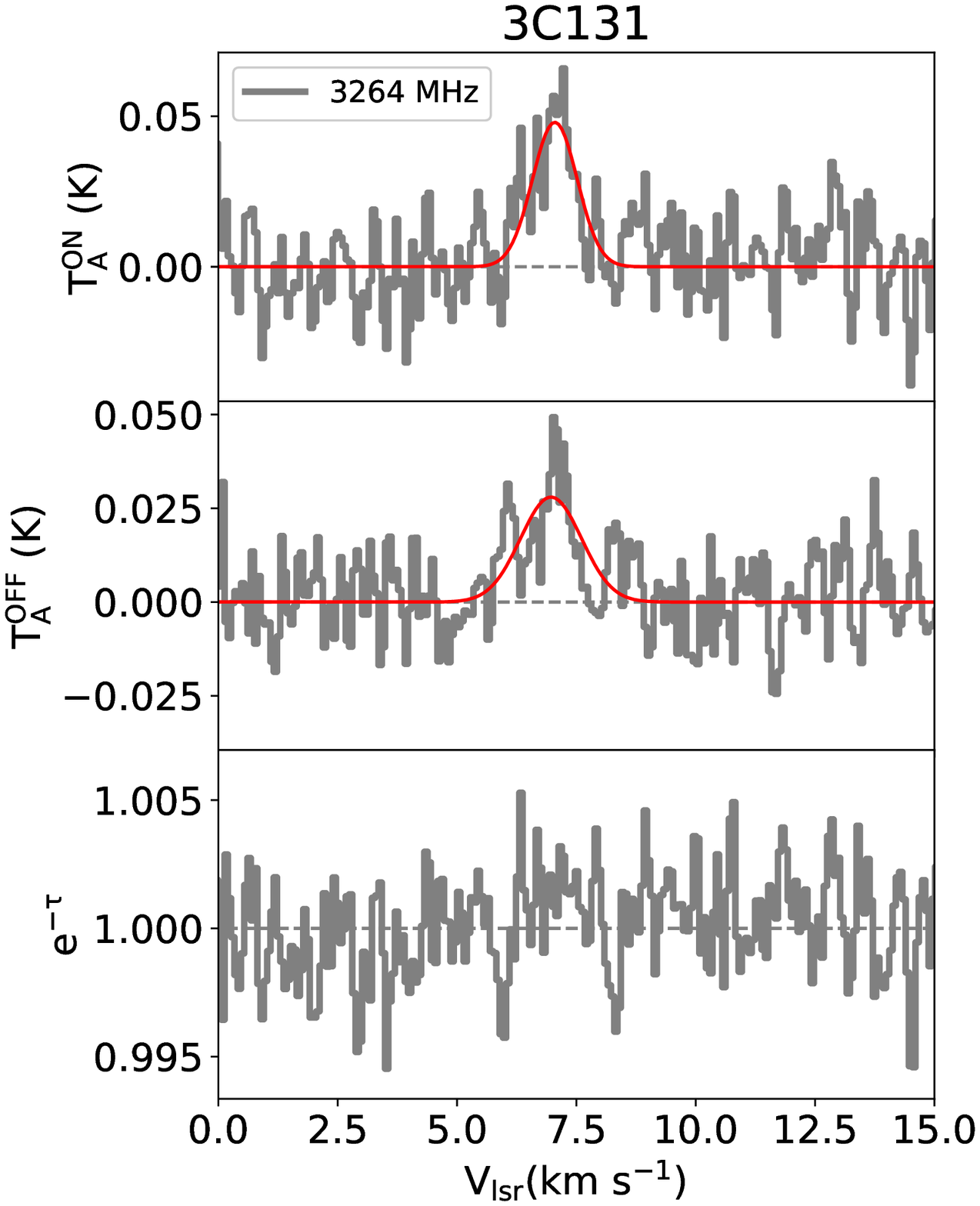}{0.32\textwidth}{}
              \fig{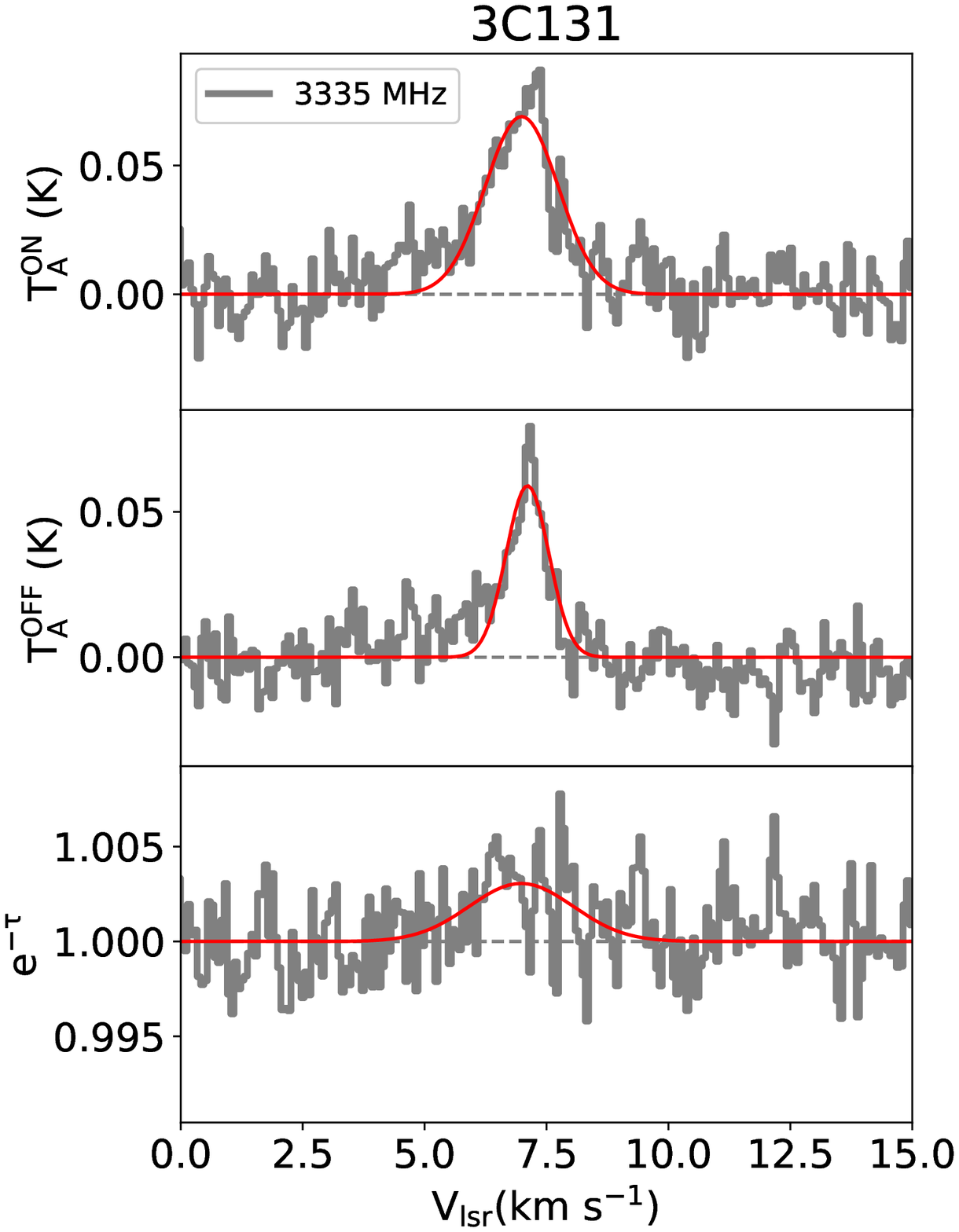}{0.32\textwidth}{}
              \fig{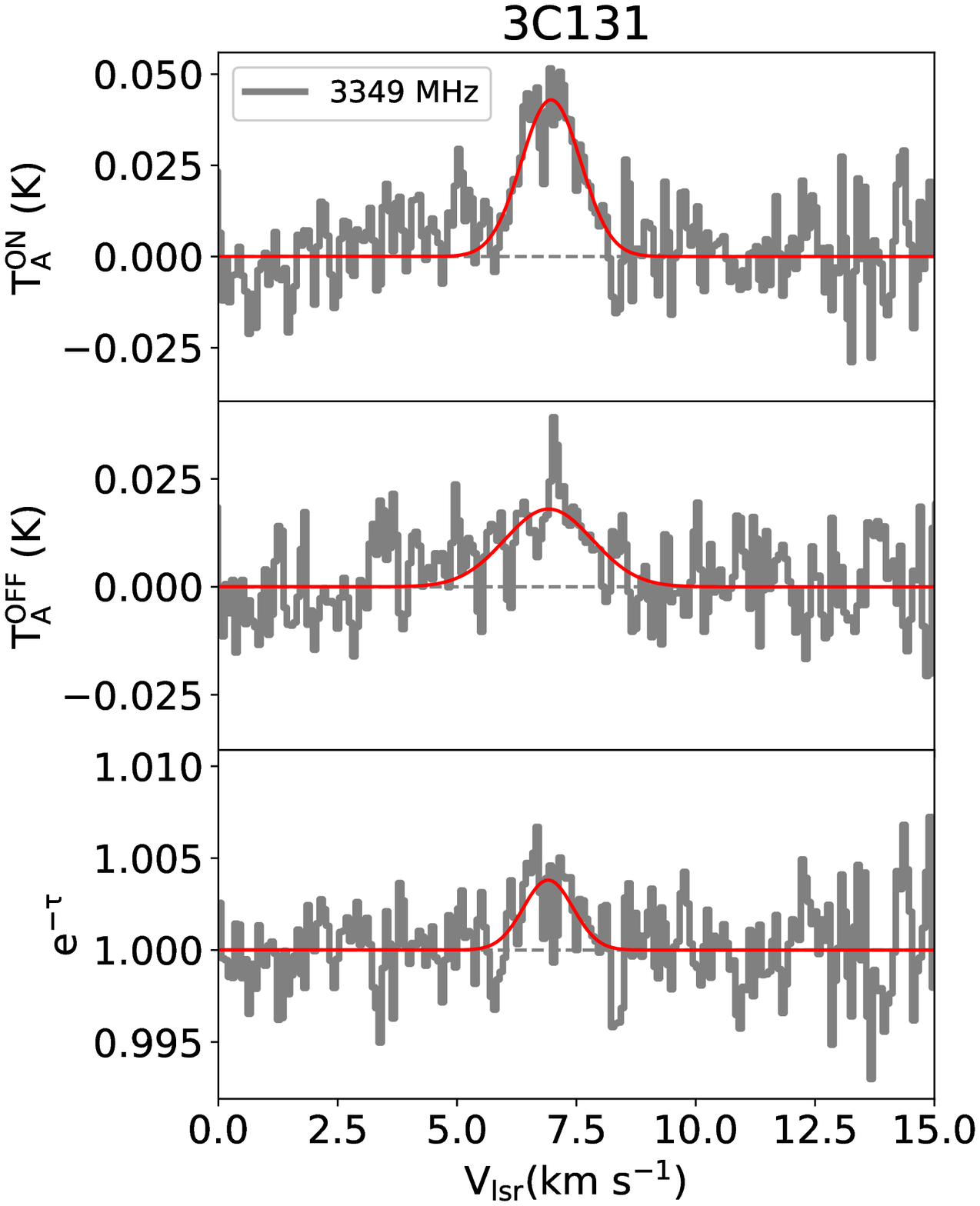}{0.32\textwidth}{}
              }
\gridline{\fig{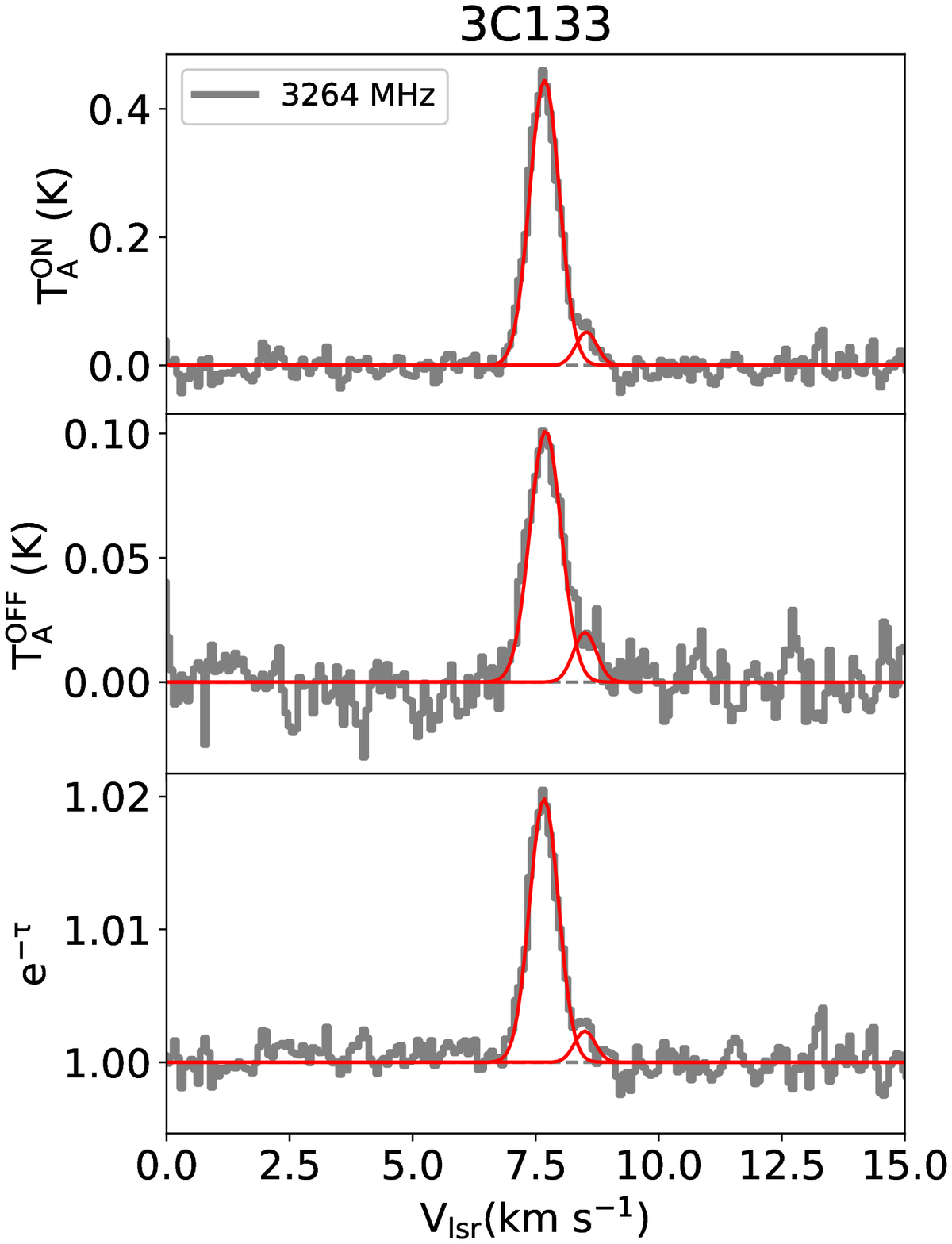}{0.32\textwidth}{}
              \fig{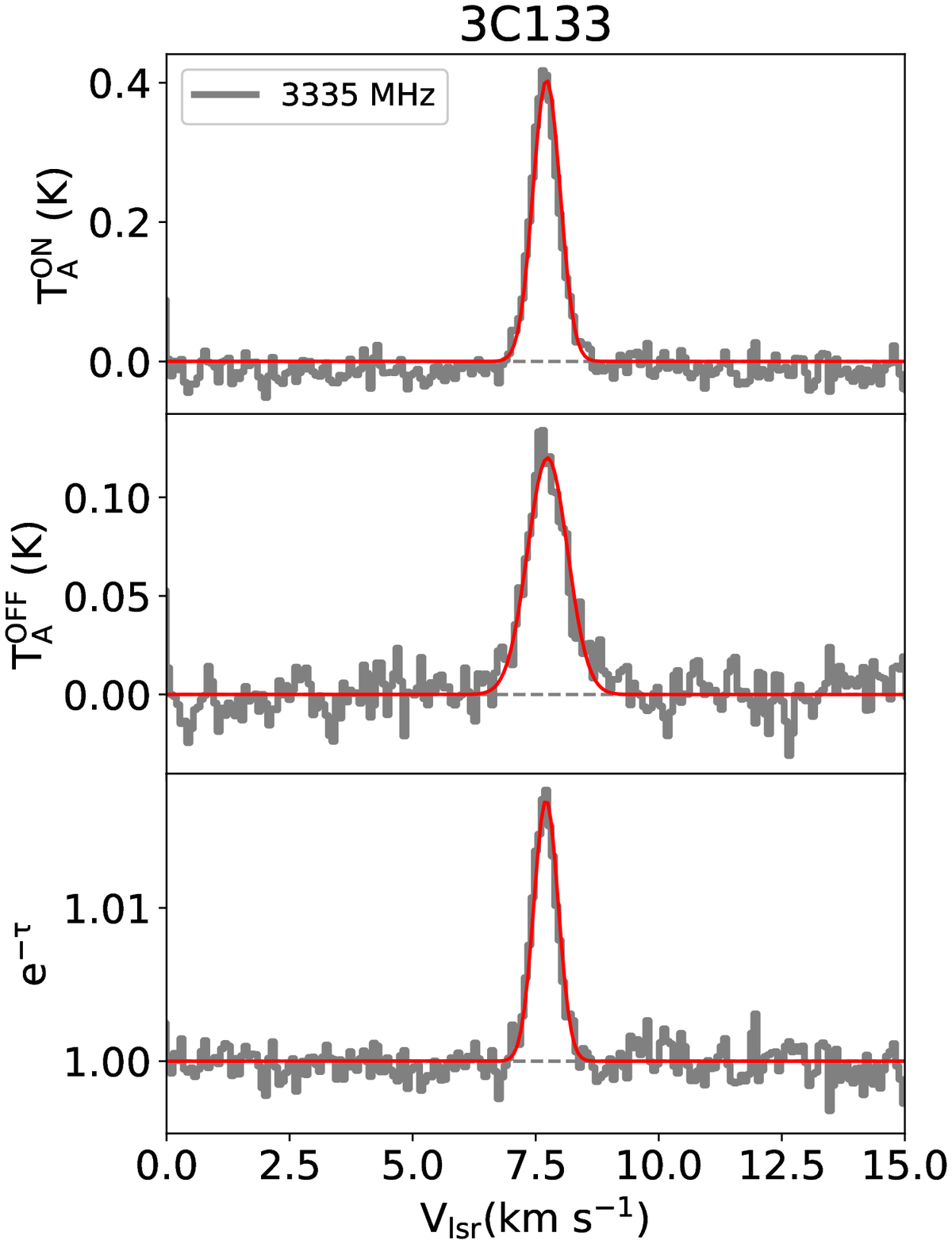}{0.32\textwidth}{}
              \fig{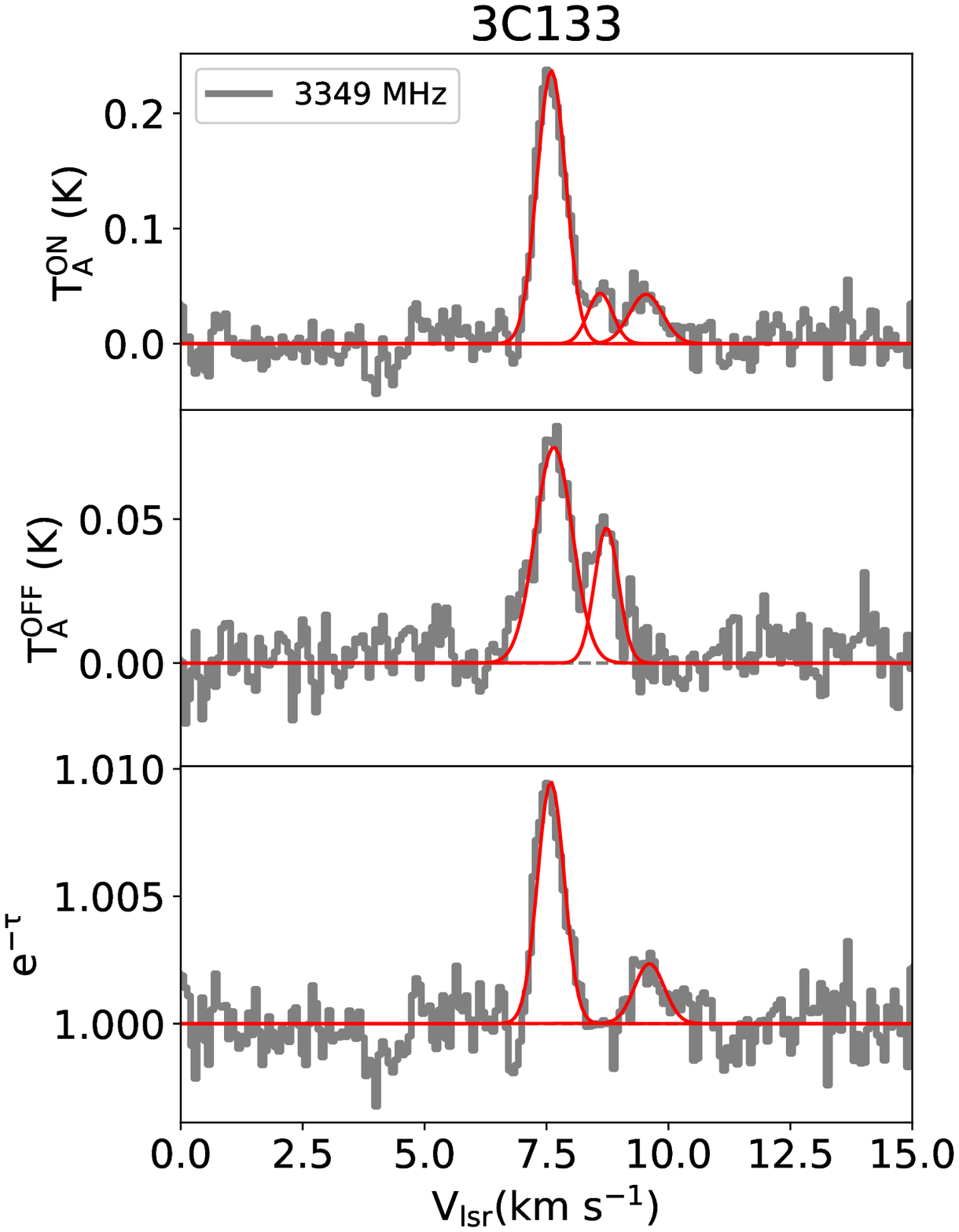}{0.32\textwidth}{}
              }
\gridline{\fig{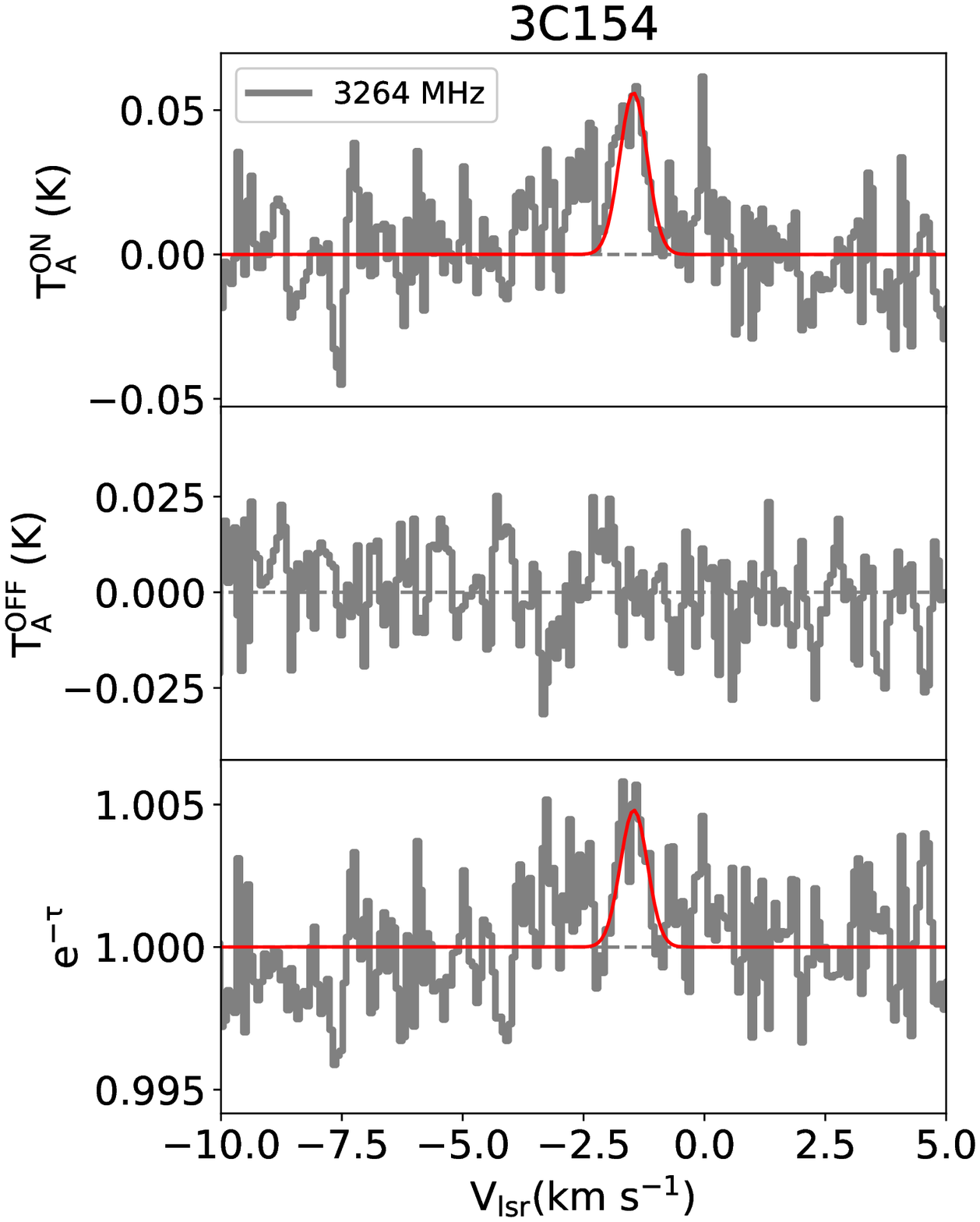}{0.33\textwidth}{}
              \fig{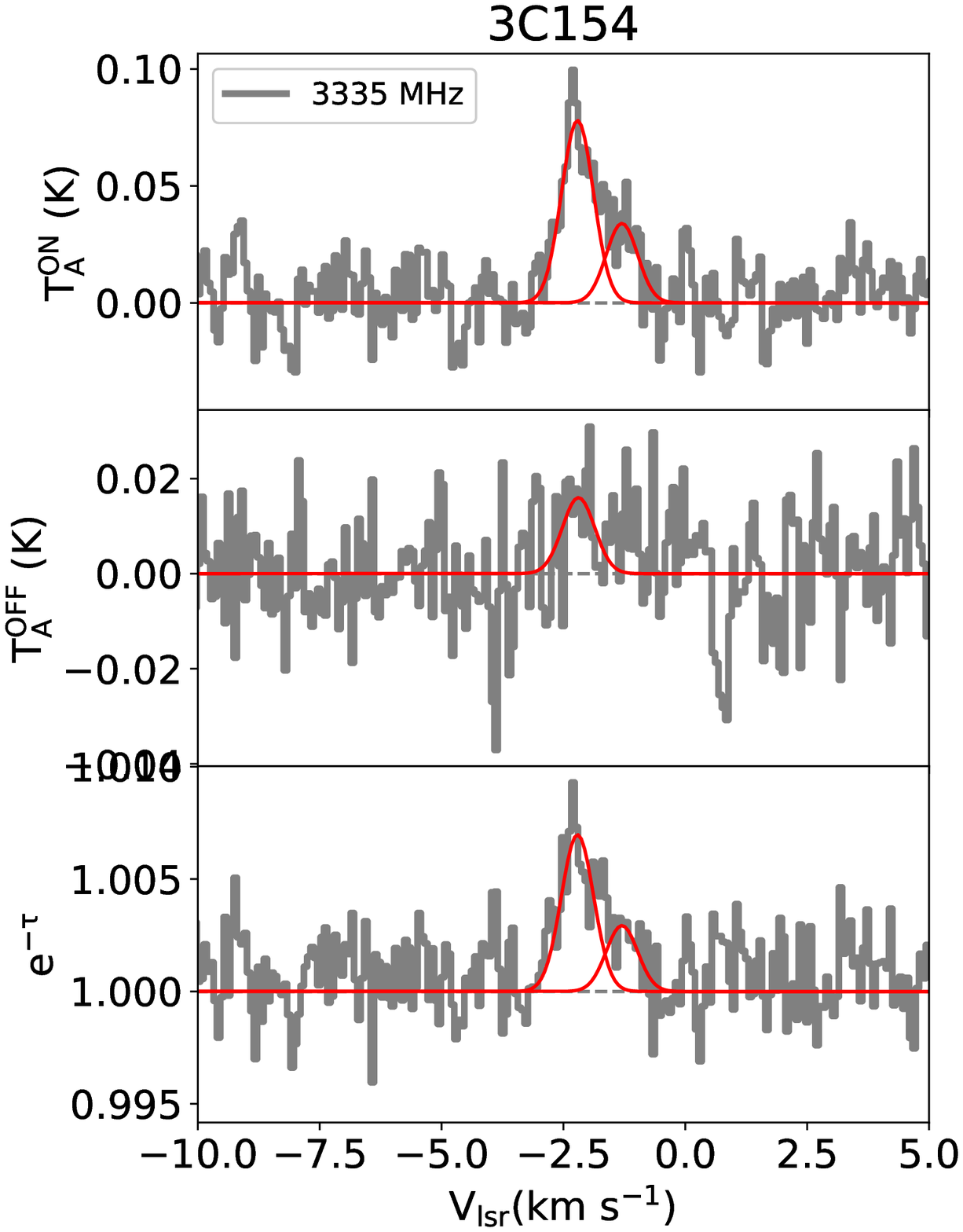}{0.33\textwidth}{}
              \fig{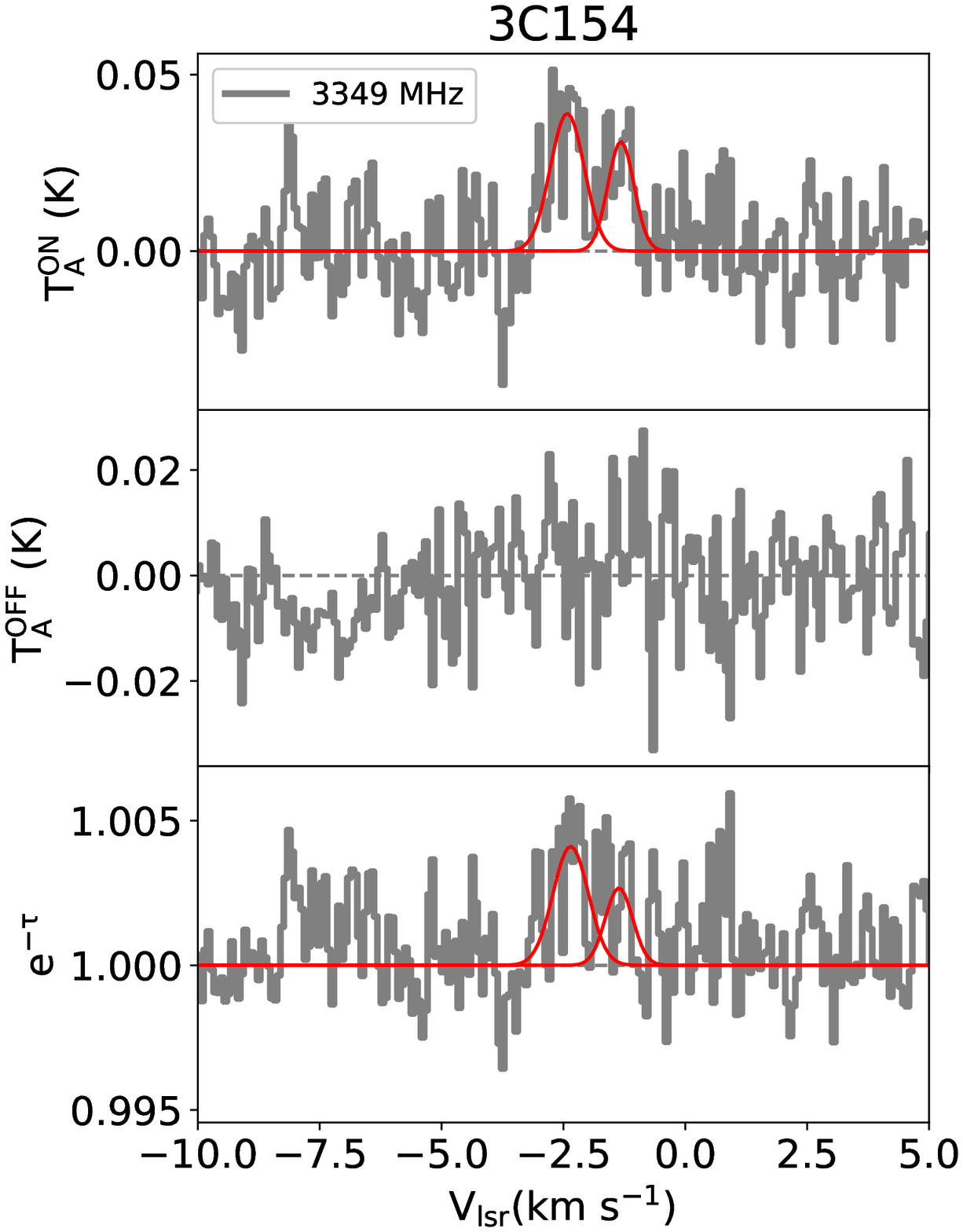}{0.33\textwidth}{}
              }
\end{figure*}

\begin{longrotatetable}
\begin{deluxetable*}{lllllllllll}
\tablecaption{Gaussian fitting parameters of CH 3335 MHz data.\label{table:chfitresult}}
\tablewidth{600pt}
\tabletypesize{\tiny}
\tablehead{
\colhead{Source} & \colhead{$l/b$} &  
\multicolumn{6}{c}{CH 3335 MHz}  \\ 
\cline{3-8}
\colhead{} & \colhead{}  &  
\colhead{$V\rm_{cen}$} & \colhead{$\Delta V$} &\colhead{$T\rm_C$ }  & \colhead{$\tau$}  &  \colhead{$T\rm_{ex}$} & \colhead{N(CH)$\rm^{LTE}_{3335}$}  & \colhead{N(CH)} & \colhead{N(OH)} \\
\colhead{} & \colhead{(\deg)}  &  
\colhead{(\kms)} & \colhead{(\kms)} & \colhead{(K)} & \colhead{($10^{-3}$)}  &  \colhead{(K)} & \colhead{(10$^{13}$\cm2)}  &  \colhead{(10$^{13}$\cm2)}  & \colhead{(10$^{14}$\cm2)} } 
\startdata
3C123 & 170.6/$-$11.7 &   4.0 (  0.2) &   1.47 ( 0.27) & 9.23e2 & $-2.9$ (0.1) & $-54.5$ ( 12.1) &  7.1(2.1) & \nodata  & 1.13(0.69) \\
3C123 & 170.6/$-$11.7 &   4.5 (  0.0) &   0.46 ( 0.03) & 9.23e2 & $-11.0$ (0.1) & $-10.5$ (  2.9) &  1.6 (0.5) & 2.1 (0.6) & 0.84(0.29) \\
3C123 & 170.6/$-$11.7 &   5.5 (  0.1) &  0.94 ( 0.08) &   9.23e2 & $-4.5$ (0.1) & $-33.2$ (  6.8) & 4.2(0.9) & 5.6 (1.3) &  1.38(0.37) \\
3C131 & 171.4/$-$7.8 &   7.0 (  0.0)  &  1.76 ( 0.10) & 5.88 & $-3.1$ (0.5) & $-38.9$ ( 10.3) &  6.3(2.0) & 8.4(2.7) & 0.96(0.49) \\
3C133 & 177.7/$-$9.9 &   7.7 (  0.0)  &  0.65 ( 0.02) & 1.72e1& $-16.9$ (0.5) & $-13.3$ (  2.4) & 4.4(0.8) & 5.8(1.1) & 0.94(1.30) \\
3C154 & 185.6/4.0 &  $-2.2$ (  0.1)   &  0.77 ( 0.12) & 9.89 &  $-7.0$ (0.8) &  $-0.4$ (  3.3) &  0.1(0.5) & \nodata &  0.19(0.12) \\
3C154 & 185.6/4.0 &  $-1.3$ (  0.1)  &  0.76 ( 0.27) & 9.89& $-0.7$< $\tau$ <0  & $-92.6$< $T\rm_{ex}$ < 0 & < 1.5 & \nodata & 0.18(0.21) \\
T0526+24 & 181.4/$-$5.2 &   7.2 (  0.1) &   1.90 ( 0.23) & 5.89& $-7.5$ (1.4) & $-18.9$ (  7.0) & 8.1(3.5) & 10.8 (4.6) & 0.17(0.12)\\
T0629+10 & 201.5/0.5 &   2.8 (  0.1) &  0.73 ( 0.12) & 8.62 & $-7.1$ (0.9) &  $-6.4$ (  4.1) &   1.0 (0.7)& 1.3 (0.9) & \nodata \\
T0629+10 & 201.5/0.5 &   3.7 (  0.0) &  0.74 ( 0.05) & 8.62 & $-33.5$ (0.9) &  $-3.6$ (  1.0) &  2.7 (0.8) & 3.5 (1.0) & 0.95(0.09) \\
T0629+10 & 201.5/0.5 &   4.7 (  0.1) &  1.45 ( 0.22) & 8.62 & $-10.7$ (0.9) & $-22.5$ (  4.8) & 10.5 (2.9) & 14.0 (3.8) & 0.13(0.07)\\
T0629+10 & 201.5/0.5 &   6.0 (  0.0) & 0.58 ( 0.06) &  8.62 & $-22.7$ (3.6) &  $-0.9$ (  4.1) & 0.36 (1.63)& 0.44 (2.00) & 0.69(0.13) \\
T0629+10 & 201.5/0.5 &   6.6 (  0.2) &  1.20 ( 0.26) & 8.62 & $-10.6$ (1.1) &  $-4.2$ (  3.2) & 1.6 (1.3)  & 2.1(1.7)  & 0.45(0.10) \\
\enddata
\end{deluxetable*}
\end{longrotatetable}

\begin{longrotatetable}
\begin{deluxetable*}{lllllllllllll}
\tablecaption{Gaussian fitting parameters of CH 3264 and 3349 MHz data.\label{table:chfitresult1}}
\tablewidth{600pt}
\tabletypesize{\tiny}
\tablehead{
\colhead{Source} & \colhead{$l/b$} &  
 \multicolumn{5}{c}{CH 3264 MHz} &  \multicolumn{4}{c}{CH 3349 MHz} \\ 
\cline{3-7}
\cline{9-13}
\colhead{} & \colhead{}  &  
\colhead{$V\rm_{cen}$} & \colhead{$\Delta V$} &\colhead{$T\rm_C$ } & \colhead{$\tau$}  &  \colhead{$T\rm_{ex}$} & \colhead{} & 
\colhead{$V\rm_{cen}$} & \colhead{$\Delta V$} &\colhead{$T\rm_C$ } & \colhead{$\tau$}  &  \colhead{$T\rm_{ex}$}   \\
\colhead{} & \colhead{(\deg)}  &  
\colhead{(\kms)} & \colhead{(\kms)}  & \colhead{(K)}& \colhead{($10^{-3}$)}  &  \colhead{(K)} & \colhead{} & 
\colhead{(\kms)} & \colhead{(\kms)} & \colhead{(K)} & \colhead{($10^{-3}$)}&  \colhead{(K)} 
} 
\startdata
3C123 &   170.6/$-$11.7   & \nodata & \nodata & 1.05e2 & \nodata & \nodata &  & 3.4 (  0.1) &  0.65 ( 0.15) & 9.13e1 & -1.2 (0.2) & -43.7 ( 36.8) \\
3C123 &   170.6/$-$11.7   &  4.4 (  0.0) &  0.68 ( 0.03) & 1.05e2 &$-$7.2 (0.0) & $-$17.3 (  2.7) & &  4.3 (  0.0) &  0.67 ( 0.04) & 9.13e1  & $-$6.3 (0.2) & $-$22.0 ( 9.5) \\
3C123 &   170.6/$-$11.7   &   5.5 (  0.0) &  0.90 ( 0.07) & 1.05e2 & $-$3.9 (0.1) & $-$10.0 (  5.5)& &  5.4 (  0.0) &  0.97 ( 0.07) & 9.13e1 & $-$3.4 (0.1) & $-$32.4 ( 11.6) \\
3C131 &  171.4/$-$7.8  & 7.0 (  0.1) &  1.10 ( 0.13) & 8.24 & $-$2.4 (0.4) & $-$26.0 (  8.0) & &  7.0 (  0.1) &  1.46 ( 0.14) & 5.77& $-$3.8 (0.9) & $-$11.0 (  6.6) \\
3C133 &   177.7/$-$9.9  &  7.7 (  0.0) &  0.71 ( 0.02) & 1.79e1 & $-$19.6 (0.5) &  $-$8.5 (  1.8) &  &  7.6 (  0.0) &  0.69 ( 0.03) &1.69e1 & $-$9.5 (0.5) & $-$17.1 (  3.7) \\
3C133 &   177.7/$-$9.9  &  8.5 (  0.1) &  0.50 ( 0.12) & 1.79e1  & $-$2.3 (0.6) &  $-$7.9 ( 13.2) & & 8.6 (  0.1) &  0.60 ( 0.18) &  1.69e1  & 0.2 (0.3) & 664.7 (1034.9) \\
3C133 &   177.7/$-$9.9  &  \nodata & \nodata & 1.79e1  & \nodata & \nodata &  & 9.5 (  0.1) &  0.80 ( 0.19) & 1.69e1 & $-$0.9<$\tau$<0 & $-$68.5<$T\rm_{ex}$<0 \\
3C154 &   185.6/4.0   &  $-$1.5 (  0.0) &  0.67 ( 0.10) & 1.10e1 & $-$2.6 <$\tau$<0 & -22.8<$T\rm_{ex}$<0  &  & $-$1.3 (  0.1) &  0.63 ( 0.17) & 9.73 & $-$0.4<$\tau$<0 & $-$161.4<$T\rm_{ex}$<0 \\
3C154 & 185.6/4.0   & \nodata & \nodata & 1.10e1& \nodata & \nodata  & & $-$2.4 (  0.1) &  0.82 ( 0.16) & 9.73 & $-$1.2<$\tau$<0 & $-$51.9<$T\rm_{ex}$<0   \\
T0526+24 &  181.4/$-$5.2   & 7.2 (  0.3) &  3.24 ( 0.22) & 6.03 & $-$2.8 (1.3) & $-$27.3 ( 21.5) & &  7.3 (  0.2) &  1.90 ( 0.14) & 5.87 & $-$0.2 (1.9) & $-$701.6 (7910.1) \\
T0629+10 &   201.5/0.5  &  2.8 (  0.1) &  1.00 ( 0.22) & 1.07e1 & $-$6.8 (0.8) &  $-$5.3 (  4.4) &   &  2.7 (  0.1) &  0.83 ( 0.13) & 8.51 & $-$8.0 (0.9) &   0.9 (  2.7) \\
T0629+10 &   201.5/0.5  &  3.6 (  0.0) &  0.74 ( 0.05) & 1.07e1 & $-$36.7 (0.8) &  $-$1.9 (  0.8) & &  3.6 (  0.0) &  0.69 ( 0.04) & 8.51 & $-$25.0 (0.8) &  $-$4.7 (  1.2) \\
T0629+10 &   201.5/0.5  &  4.5 (  0.0) &  0.71 ( 0.06) & 1.07e1 & $-$20.5 (0.8) &  $-$4.4 (  1.6) & &  4.6 (  0.0) &  1.08 ( 0.10) & 8.51 & $-$16.0 (0.9) &  $-$5.3 (  1.9) \\
T0629+10 &   201.5/0.5  &  5.2 (  0.0) &  0.31 ( 0.06) & 1.07e1 & $-$6.1 (1.2) &  $-$5.3 (  6.9) &  &  6.0 (  0.0) &  0.74 ( 0.05) & 8.51 & $-$17.9 (0.8) &  $-$5.4 (  1.6) \\
T0629+10 &   201.5/0.5  &  6.1 (  0.0) &  1.21 ( 0.05) & 1.07e1 & $-$13.6 (0.6) &  $-$6.7 (  2.0) &  &  6.9 (  0.1) &  0.82 ( 0.12) & 8.51 & $-$6.6 (0.8) &  $-$7.1 (  4.0) \\
\enddata
\end{deluxetable*}
\end{longrotatetable}


\begin{thebibliography}{}

\bibitem[Allen(1994)]{1994ApJ...424..754A} Allen, M.~M.\ 1994, \apj, 424, 754. doi:10.1086/173928
\bibitem[Black \& Dalgarno(1973)]{1973ApL....15...79B} Black, J.~H. \& Dalgarno, A.\ 1973, \aplett, 15, 79
\bibitem[Black et al.(1975)]{1975ApJ...199..633B} Black, J.~H., Dalgarno, A., \& Oppenheimer, M.\ 1975, \apj, 199, 633. doi:10.1086/153730
\bibitem[Bohlin et al.(1978)]{1978ApJ...224..132B} Bohlin, R.~C., Savage, B.~D., \& Drake, J.~F.\ 1978, \apj, 224, 132. doi:10.1086/156357
\bibitem[Boiss{\'e} et al.(2005)]{2005A&A...429..509B} Boiss{\'e}, P., Le Petit, F., Rollinde, E., et al.\ 2005, \aap, 429, 509. doi:10.1051/0004-6361:20047135
\bibitem[Crane et al.(1995)]{1995ApJS...99..107C} Crane, P., Lambert, D.~L., \& Sheffer, Y.\ 1995, \apjs, 99, 107. doi:10.1086/192180
\bibitem[Crawford(1995)]{1995MNRAS.277..458C} Crawford, I.~A.\ 1995, \mnras, 277, 458. doi:10.1093/mnras/277.2.458
\bibitem[Crutcher(1985)]{1985ApJ...288..604C} Crutcher, R.~M.\ 1985, \apj, 288, 604. doi:10.1086/162826
\bibitem[Dailey et al.(2020)]{2020MNRAS.495..510D} Dailey, E.~M., Smith, A.~J., Magnani, L., et al.\ 2020, \mnras, 495, 510. doi:10.1093/mnras/staa1188
\bibitem[Danks et al.(1984)]{1984A&A...130...62D} Danks, A.~C., Federman, S.~R., \& Lambert, D.~L.\ 1984, \aap, 130, 62
\bibitem[Dickey et al.(1983)]{1983ApJS...53..591D} Dickey, J.~M., Kulkarni, S.~R., van Gorkom, J.~H., et al.\ 1983, \apjs, 53, 591. doi:10.1086/190903
\bibitem[Dickey et al.(1981)]{1981A&A....98..271D} Dickey, J.~M., Crovisier, J., \& Kazes, I.\ 1981, \aap, 98, 271
\bibitem[Draine(1986)]{1986ApJ...310..408D} Draine, B.~T.\ 1986, \apj, 310, 408. doi:10.1086/164694
\bibitem[Draine \& Katz(1986a)]{1986ApJ...306..655D} Draine, B.~T. \& Katz, N.\ 1986, \apj, 306, 655. doi:10.1086/164375
\bibitem[Draine \& Katz(1986b)]{1986ApJ...310..392D} Draine, B.~T. \& Katz, N.\ 1986, \apj, 310, 392. doi:10.1086/164693
\bibitem[Dunham(1937)]{1937PASP...49...26D} Dunham, T.\ 1937, \pasp, 49, 26. doi:10.1086/124759
\bibitem[Elitzur \& Watson(1978)]{1978ApJ...222L.141E} Elitzur, M. \& Watson, W.~D.\ 1978, \apjl, 222, L141. doi:10.1086/182711
\bibitem[Federman(1982)]{1982ApJ...257..125F} Federman, S.~R.\ 1982, \apj, 257, 125. doi:10.1086/159970
\bibitem[Federman et al.(1994)]{1994ApJ...424..772F} Federman, S.~R., Strom, C.~J., Lambert, D.~L., et al.\ 1994, \apj, 424, 772. doi:10.1086/173930
\bibitem[Federman et al.(1996)]{1996MNRAS.279L..41F} Federman, S.~R., Rawlings, J.~M.~C., Taylor, S.~D., et al.\ 1996, \mnras, 279, L41. doi:10.1093/mnras/279.3.L41
\bibitem[Felenbok \& Roueff(1996)]{1996ApJ...465L..57F} Felenbok, P. \& Roueff, E.\ 1996, \apjl, 465, L57. doi:10.1086/310129
\bibitem[Flower \& Pineau des Forets(1998)]{1998MNRAS.297.1182F} Flower, D.~R. \& Pineau des Forets, G.\ 1998, \mnras, 297, 1182. doi:10.1046/j.1365-8711.1998.01574.x
\bibitem[Genzel et al.(1979)]{1979A&A....73..253G} Genzel, R., Downes, D., Pauls, T., et al.\ 1979, \aap, 73, 253
\bibitem[Gredel(1997)]{1997A&A...320..929G} Gredel, R.\ 1997, \aap, 320, 929
\bibitem[Gredel et al.(1993)]{1993A&A...269..477G} Gredel, R., van Dishoeck, E.~F., \& Black, J.~H.\ 1993, \aap, 269, 477
\bibitem[Grenier et al.(2005)]{2005Sci...307.1292G} Grenier, I.~A., Casandjian, J.-M., \& Terrier, R.\ 2005, Science, 307, 1292. doi:10.1126/science.1106924
\bibitem[Hawkins \& Craig(1991)]{1991ApJ...375..642H} Hawkins, I. \& Craig, N.\ 1991, \apj, 375, 642. doi:10.1086/170227
\bibitem[Heiles \& Troland(2003a)]{2003ApJS..145..329H} Heiles, C. \& Troland, T.~H.\ 2003, \apjs, 145, 329. doi:10.1086/367785
\bibitem[Heiles \& Troland(2003b)]{2003ApJ...586.1067H} Heiles, C. \& Troland, T.~H.\ 2003, \apj, 586, 1067. doi:10.1086/367828
\bibitem[Hjalmarson et al.(1977)]{1977ApJS...35..263H} Hjalmarson, A., Sume, A., Elider, J., et al.\ 1977, \apjs, 35, 263. doi:10.1086/190480
\bibitem[Jacq et al.(1987)]{1987A&A...173..347J} Jacq, T., Baudry, A., Despois, D., et al.\ 1987, \aap, 173, 347
\bibitem[Jacob et al.(2021)]{2021A&A...650A.133J} Jacob, A.~M., Menten, K.~M., Wiesemeyer, H., et al.\ 2021, \aap, 650, A133. doi:10.1051/0004-6361/202140419
\bibitem[Jenniskens et al.(1992)]{1992A&A...265L...1J} Jenniskens, P., Ehrenfreund, P., \& Desert, F.-X.\ 1992, \aap, 265, L1
\bibitem[Lang \& Wilson(1978)]{1978ApJ...224..125L} Lang, K.~R. \& Wilson, R.~F.\ 1978, \apj, 224, 125. doi:10.1086/156356
\bibitem[Li et al.(2015)]{2015PKAS...30...75L} Li, D., Xu, D., Heiles, C., et al.\ 2015, Publication of Korean Astronomical Society, 30, 75. doi:10.5303/PKAS.2015.30.2.075
\bibitem[Li et al.(2018a)]{2018ApJS..235....1L} Li, D., Tang, N., Nguyen, H., et al.\ 2018a, \apjs, 235, 1. doi:10.3847/1538-4365/aaa762
\bibitem[Li et al.(2018b)]{2018IMMag..19..112L} Li, D., Wang, P., Qian, L., et al.\ 2018b, IEEE Microwave Magazine, 19, 112. doi:10.1109/MMM.2018.2802178
\bibitem[Li et al.(2019)]{2019RAA....19...16L} Li, D., Dickey, J.~M., \& Liu, S.\ 2019, Research in Astronomy and Astrophysics, 19, 016. doi:10.1088/1674-4527/19/2/16
\bibitem[Liszt \& Lucas(2002)]{2002A&A...391..693L} Liszt, H. \& Lucas, R.\ 2002, \aap, 391, 693. doi:10.1051/0004-6361:20020849
\bibitem[Litvak(1969)]{1969ApJ...156..471L} Litvak, M.~M.\ 1969, \apj, 156, 471. doi:10.1086/149982
\bibitem[Lucas \& Liszt(1996)]{1996A&A...307..237L} Lucas, R. \& Liszt, H.\ 1996, \aap, 307, 237
\bibitem[Luo et al.(2020)]{2020ApJ...889L...4L} Luo, G., Li, D., Tang, N., et al.\ 2020, \apjl, 889, L4. doi:10.3847/2041-8213/ab6337
\bibitem[Magnani et al.(1992)]{1992A&AS...93..509M} Magnani, L., Sandell, G., \& Lada, E.~A.\ 1992, \aaps, 93, 509
\bibitem[Magnani \& Onello(1993)]{1993ApJ...408..559M} Magnani, L. \& Onello, J.~S.\ 1993, \apj, 408, 559. doi:10.1086/172613
\bibitem[Magnani et al.(1998)]{1998ApJ...504..290M} Magnani, L., Onello, J.~S., Adams, N.~G., et al.\ 1998, \apj, 504, 290. doi:10.1086/306062
\bibitem[Mattila(1986)]{1986A&A...160..157M} Mattila, K.\ 1986, \aap, 160, 157
\bibitem[McKellar(1940)]{1940PASP...52..187M} McKellar, A.\ 1940, \pasp, 52, 187. doi:10.1086/125159
\bibitem[Nguyen et al.(2018)]{2018ApJ...862...49N} Nguyen, H., Dawson, J.~R., Miville-Desch{\^e}nes, M.-A., et al.\ 2018, \apj, 862, 49. doi:10.3847/1538-4357/aac82b
\bibitem[Penprase(1993)]{1993ApJS...88..433P} Penprase, B.~E.\ 1993, \apjs, 88, 433. doi:10.1086/191829
\bibitem[Rachford et al.(2001)]{2001ApJ...555..839R} Rachford, B.~L., Snow, T.~P., Tumlinson, J., et al.\ 2001, \apj, 555, 839. doi:10.1086/321489
\bibitem[Roueff(1996)]{1996MNRAS.279L..37R} Roueff, E.\ 1996, \mnras, 279, L37. doi:10.1093/mnras/279.3.L37
\bibitem[Rydbeck et al.(1976)]{1976ApJS...31..333R} Rydbeck, O.~E.~H., Kollberg, E., Hjalmarson, A., et al.\ 1976, \apjs, 31, 333. doi:10.1086/190385
\bibitem[Sakai et al.(2012)]{2012A&A...546A.103S} Sakai, N., Maezawa, H., Sakai, T., et al.\ 2012, \aap, 546, A103. doi:10.1051/0004-6361/201219106
\bibitem[Sandell et al.(1981)]{1981A&A....97..317S} Sandell, G., Johansson, L.~E.~B., Rieu, N.~Q., et al.\ 1981, \aap, 97, 317
\bibitem[Sandell et al.(1980)]{1980A&A....83..226S} Sandell, G., Hoglund, B., \& Friberg, P.\ 1980, \aap, 83, 226
\bibitem[Sandell et al.(1988)]{1988ApJ...329..920S} Sandell, G., Magnani, L., \& Lada, E.~A.\ 1988, \apj, 329, 920. doi:10.1086/166436
\bibitem[Savage et al.(1977)]{1977ApJ...216..291S} Savage, B.~D., Bohlin, R.~C., Drake, J.~F., et al.\ 1977, \apj, 216, 291. doi:10.1086/155471
\bibitem[Schlegel et al.(1998)]{1998ApJ...500..525S} Schlegel, D.~J., Finkbeiner, D.~P., \& Davis, M.\ 1998, \apj, 500, 525. doi:10.1086/305772
\bibitem[Sheffer et al.(2008)]{2008ApJ...687.1075S} Sheffer, Y., Rogers, M., Federman, S.~R., et al.\ 2008, \apj, 687, 1075. doi:10.1086/591484
\bibitem[Suutarinen et al.(2011)]{2011A&A...531A.121S} Suutarinen, A., Geppert, W.~D., Harju, J., et al.\ 2011, \aap, 531, A121. doi:10.1051/0004-6361/201016079
\bibitem[Swings \& Rosenfeld(1937)]{1937ApJ....86..483S} Swings, P. \& Rosenfeld, L.\ 1937, \apj, 86, 483. doi:10.1086/143880
\bibitem[Planck Collaboration et al.(2011)]{2011A&A...536A..19P} Planck Collaboration, Ade, P.~A.~R., Aghanim, N., et al.\ 2011, \aap, 536, A19. doi:10.1051/0004-6361/201116479
\bibitem[van Dishoeck \& Black(1988)]{1988ApJ...334..771V} van Dishoeck, E.~F. \& Black, J.~H.\ 1988, \apj, 334, 771. doi:10.1086/166877
\bibitem[van Dishoeck \& Black(1989)]{1989ApJ...340..273V} van Dishoeck, E.~F. \& Black, J.~H.\ 1989, \apj, 340, 273. doi:10.1086/167391
\bibitem[Weselak et al.(2010)]{2010MNRAS.402.1991W} Weselak, T., Galazutdinov, G.~A., Beletsky, Y., et al.\ 2010, \mnras, 402, 1991. doi:10.1111/j.1365-2966.2009.16028.x
\bibitem[Weselak et al.(2009)]{2009A&A...499..783W} Weselak, T., Galazutdinov, G., Beletsky, Y., et al.\ 2009, \aap, 499, 783. doi:10.1051/0004-6361/200911616
\bibitem[Xu \& Li(2016)]{2016ApJ...833...90X} Xu, D. \& Li, D.\ 2016, \apj, 833, 90. doi:10.3847/1538-4357/833/1/90
\bibitem[Zuckerman \& Turner(1975)]{1975ApJ...197..123Z} Zuckerman, B. \& Turner, B.~E.\ 1975, \apj, 197, 123. doi:10.1086/153492
%
%
\end{thebibliography}
\end{document}